\documentclass[a4paper,11pt]{article}
\usepackage{jcappub} 
                     
\usepackage{url}
\usepackage{hyperref}
\usepackage{latexsym}
\usepackage{amssymb}
\usepackage{graphicx}
\usepackage{color}

\usepackage{mathptmx}       
\usepackage{helvet}         
\usepackage{courier}        
\usepackage{type1cm}        
\usepackage{makeidx}         
\usepackage{multicol}        
\usepackage[bottom]{footmisc}

\title{Low redshift observational constraints on tachyon models of dark energy}

\author[a]{Avinash Singh}
\author[a]{Archana Sangwan}
\author[a]{H. K. Jassal}

\affiliation[a]{Indian Institute of Science Education and Research
Mohali,\\ SAS Nagar, Mohali 140306, Punjab, India.}

\emailAdd{avinashsingh@iisermohali.ac.in}
\emailAdd{archanakumari@iisermohali.ac.in}
\emailAdd{hkjassal@iisermohali.ac.in}

\abstract{The background evolution of an accelerated, dark energy
  dominated universe is aptly described by non-canonical tachyon 
  scalar field  models.
  The accelerated expansion of the universe is determined by the choice
  of a suitable scalar field potential; in the case of a tachyon
  field, a 'runaway' potential.
In the absence of a fundamental theory, dark energy properties are studied in a
phenomenological approach. 
This includes determining the model parameters using observations and
to probe the allowed deviation from the cosmological constant model.
In this paper, we present constraints on tachyon scalar field
parameters from low redshift data for two different scalar
field potentials.
These scalar field potentials have been crucial in tachyon dark energy studies.
The datasets considered in this paper include the supernova type Ia data, independent
measurements of the Hubble parameter and the Baryon Acoustic
Oscillation data.
In this paper, we present constraints on tachyon field parameters using
these observations and their combination.
A combination of the datasets indicates that those model parameters
are preferred which emulate the cosmological
constant model.
The initial value of the scalar field, in the unit of the Hubble constant,
is bounded from below and does not require fine-tuning at larger values.
}

\makeindex

\begin{document}
\maketitle
\flushbottom
\section{Introduction}
    \label{sec::intro}
Observations have established that more than two-thirds of the energy density
of the universe is due to the contribution of dark
energy~\cite{Ade:2013sjv,Planck-2015,Planck-2018}. 
Dark energy accounts for the observed late-time acceleration of the   
universe~\cite{Perlmutter1997,Perlmutter1999,Riess1998}. 
The nature of dark energy is, as yet, a mystery. 
To understand the nature of dark energy many models have been purposed,
the simplest and the most favoured  being the cosmological constant
model~\cite{Carroll1992,Carroll:2000fy}. 
For this component, the energy density remains a constant and the
equation of state parameter is given by $w=-1$. 
Although this model is consistent with cosmological 
observations, the attempts to explain the cosmological constant as the
energy density of vacuum suffers from the fine-tuning
problem~\cite{Carroll:2000fy}. 
While the theoretical problem of fine-tuning remains, the cosmological
constant model is the concordant model of dark energy. 

Observations do not rule out models with an equation of state
with $w \ne -1$, which is the property of other models  such as the barotropic
fluid models, canonical scalar field models, non-canonical scalar field models
etc.
The dynamical nature of the dark energy equation of state parameter is
assumed by considering a functional form or parametrization of $w$. 
These parameterizations include those in which the equation of state
parameter is  a constant or is a function of time. 
The two key parameters are the present day value of the equation of state parameter and its 
derivatives.
The simplest and most widely used prescription to model varying dark energy is the Chevallier-Polarski-Linder (CPL) parameterization \cite{Chevallier2001, Linder2003}.
This function has been widely employed in theoretical and observational studies of dark energy.
Many other parameterizations have been described in~\cite{JBP2005,Efstathiou1999,Lee2005,Hannestad2004,WangTegmark2004,
Huterer2001}.
The parameters are then constrained using different datasets, a few
examples of such work
are~\cite{Tripathi:2016slv,Sangwan:2017kxi,ZHENG2017,KUMAR2014,Rezaei2017,Yang2018,caozhu2014}. 

Scalar fields are a well studied class of models for dark energy where
the late time acceleration is achieved as the field evolves. 
The scalar field models include the canonical scalar
fields such as quintessence fields and the non-canonical scalar fields
such as tachyon fields and K-essence fields being  a few examples.
In  scalar field models, the equation of state depends upon
the functional form of the scalar field potential and on whether the
kinetic energy term is sub-dominant or is the driving component of the
evolution of the universe.
In order to have an accelerated expansion, a slow rolling field is required.
Although the scalar field models alleviate the fine-tuning problem
that $\Lambda$CDM model suffers from, to obtain the accelerated
expansion in the recent era, they require tuning of their own.     
A detailed study for quintessence scalar field model is given
in~\cite{Ratra:1987rm, 
Linder:2007wa,Huterer:2006mv,Zlatev:1998tr,Copeland:1997et,Sangwan:2018zpz,
Watson:2003kk,Scherrer:2007pu,Dutta:2011ik,ChenCaoetal2015}. 

Tachyon scalar field belongs to the class of non-canonical scalar field
models and has been studied in detail in \cite{BJP2003}.
Tachyon field arises naturally in string theory as a decay mode of
D-branes~\cite{asen2002a,asen2002b,asen2002c}. 
This field is a viable model of cosmology, and it has been shown that the
tachyon scalar field can effectively explain dark energy. 
Apart from explaining accelerated expansion, an important property of
this model is that its equation of state becomes dust like at early
times, i.e., the equation of state parameter becomes zero in
the past~\cite{BJP2003,Paddy2002,Calcagni2006}. 
The dynamics of the universe in this case are driven by a 'runaway'
potential~\cite{Copeland2005,Aguirregabiria2004}. 
The dark energy equation of state parameter is limited to the range
$-1 \le w_\phi \le 0$ and hence phantom like equation of state is ruled
out in this case.    
These considerations make this description of dark energy an
interesting alternative to both fluid and canonical scalar field
models. 
The tachyon models have also been extensively studied in the context
of inflation~\cite{FAIRBAIRN2002,KofmanLinde2002,Feinstein2002,
RezazadehKaramiHashemi2017,Nandinii_et_all2015,Fei_et_al2017,GaoGongFei2018,Nndinii_et_al2018}. 
Since tachyon scalar field shows dust like equation of state in its
cosmic evolution, the 'tachyon dust' is considered as a potential
candidate of combined dark energy and dark matter ~\cite{PadyChoudhury2002,asen2002b,
asen2002c,SugimotoTerashima2002,Das2004,Davies2004,Makukov2016}.

In this paper, we assume that the dark energy is described by a
homogeneous tachyon field. 
We consider two models; one with an inverse power law potential and
another with an exponential potential which have been the default
potentials being used  for studying tachyon field.
We revisit the constraints on tachyon dark energy model with new data
set of Baryon Acoustic Oscillations (BAO)~\cite{Seo2003,Percival2007,sca2013,Blake2012, 
Anderson2014,Veropalumbo2014,Delubac2015}, Supernova Type Ia (SN-Ia)
\cite{Suzuki2012,Perlmutter1997,Perlmutter1999,Riess1998,Garnavich1998,Tonry2003,Barris2004, 
Goobar2000,Gonzalez2012,Astier2006} and direct measurements of Hubble parameter (H(z))
\cite{Farooq2017,Farooq2013a,Farooq2013b,Samushia2006,Stern2010,Moresco2012,Chuang2013,Chen2011}.
We obtain stringent constraints on tachyon field parameters, by way of
combining these datasets.
Our motivation is to compare the constraints on the tachyon models
from previous studies using the same datasets and to check if the
non-canonical scalar field models prefer a different combination(s) of
cosmological parameters.
In this analysis, we have restricted ourselves to the low redshift datasets.
The structure of the paper is as follows.
In the next section~\ref{sec::BGcosmology}, we discuss  the background
cosmology in the presence of a tachyon field and  two different
scalar potentials. 
The different low redshift observational datasets used to constrain
the model parameters are  discussed in section~\ref{sec::observation}.    
In section~\ref{sec::results}, we discuss the results of the analysis
for the models. 
We summarise and conclude the paper in section~\ref{sec::SummaryandConclusions}. 
\section{Solutions of Cosmological Equations}
  \label{sec::BGcosmology}
The dynamics of the universe is governed by the Friedman equations which
are given by
    \begin{equation}
      \label{eq:friedmann}
    	\left( \frac{\dot{a}}{a} \right )^2 = \frac{8\pi G}{3}\rho 
    	 \, \,,\quad \frac{\ddot{a}}{a} = -\frac{4\pi G}{3}(\rho + 3P),
    \end{equation}
where $\rho = \rho_m + \rho_r + \rho_\phi$.
The quantities $\rho_m$ and $\rho_r$ are energy densities of
non-relativistic matter(baryonic matter + dark matter ) and
relativistic matter respectively whereas $\rho_\phi$ represents
energy density of the tachyon field.
The tachyon scalar field is described by the Lagrangian 
 \begin{equation}
	L = -V(\phi)\sqrt{1-\partial^\mu \phi \partial_\mu \phi},
 \end{equation}
where $V(\phi)$ is an arbitrary potential.
The energy density and pressure of tachyon field are
\begin{equation}
	\label{eq:energypressure}
	\rho_\phi = \frac{V(\phi)}{\sqrt{1-\dot{\phi}^2}} \, \,,\quad P_\phi = -V(\phi)\sqrt{1-\dot{\phi}^2}.
 \end{equation}  
Therefore, the equation of state parameter of the tachyon field is
 $w_{\phi} = P_\phi / \rho_\phi = \dot{\phi}^2 -1 $. 
The dynamics of the scalar field is governed by the equation of motion for the scalar field
 \begin{equation}
    \label{eq:scalar_dynamics}
	\ddot{\phi} = -(1-\dot{\phi}^2) \left [ 3H\dot{\phi} + \frac{1}{V(\phi)}\frac{dV}{d\phi}\right].
 \end{equation}
As $\dot{\phi}$ approaches $\pm 1$, the equation of state becomes dust
like, and the quantity $\ddot{\phi}$ goes to zero.
Therefore, the equation of state  remains dust like for a long time.
The cosmological evolution in this model depends on the choice of 
potential.
We consider two runaway potentials which have been employed to study
tachyon dynamics. The runaway potential naturally arises in string theory and
M-theory, and they are capable of generating the late time accelerated
expansion of the
universe~\cite{Barreiro1998,Binétruy1999,ROSATI20035,Ngampitipan2011}. 
The background cosmology in the presence of two different tachyon
scalar field potentials is summarized below. 
\begin{figure}[t]
        \centering
        \includegraphics[width=0.32\textwidth,angle=-90]{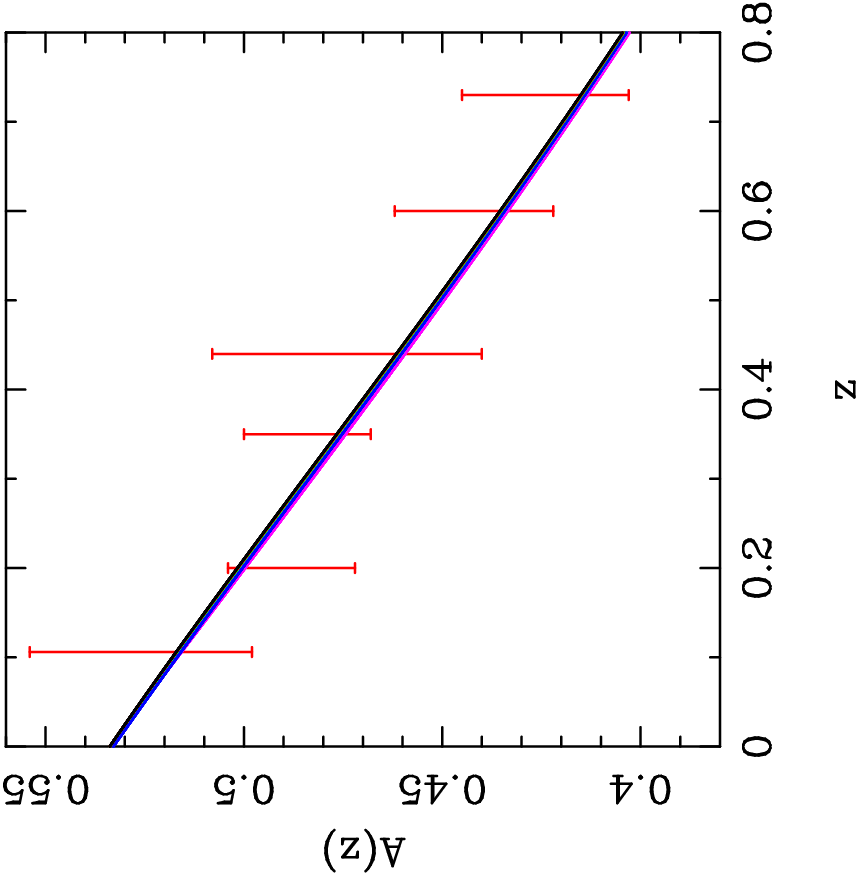}
        \includegraphics[width=0.32\textwidth,angle=-90]{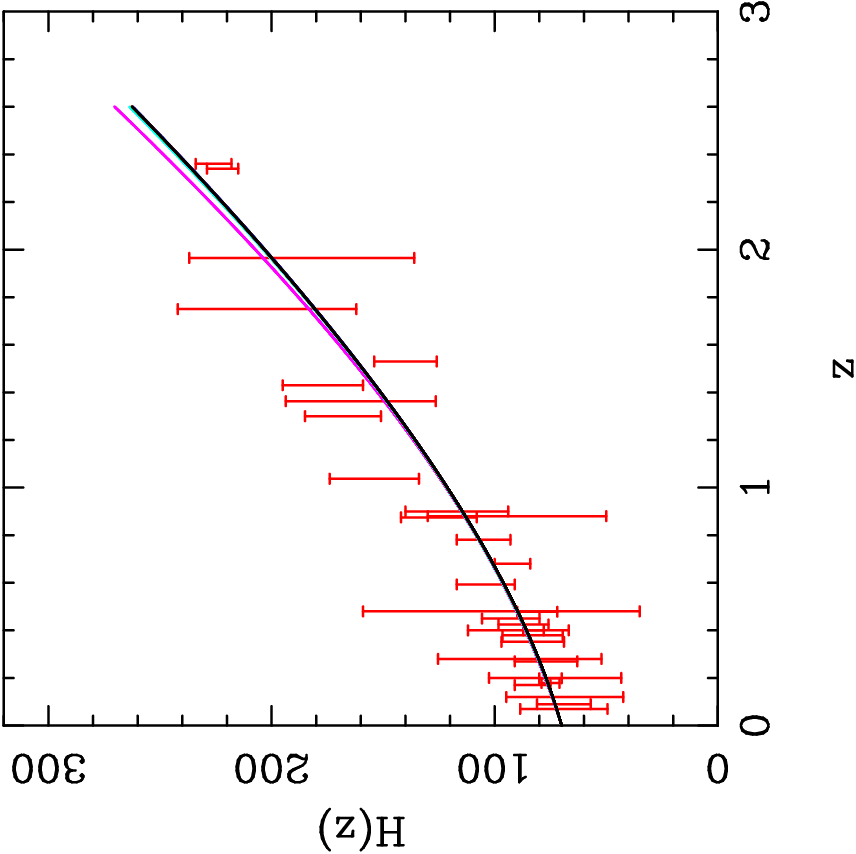}
        \includegraphics[width=0.32\textwidth,angle=-90]{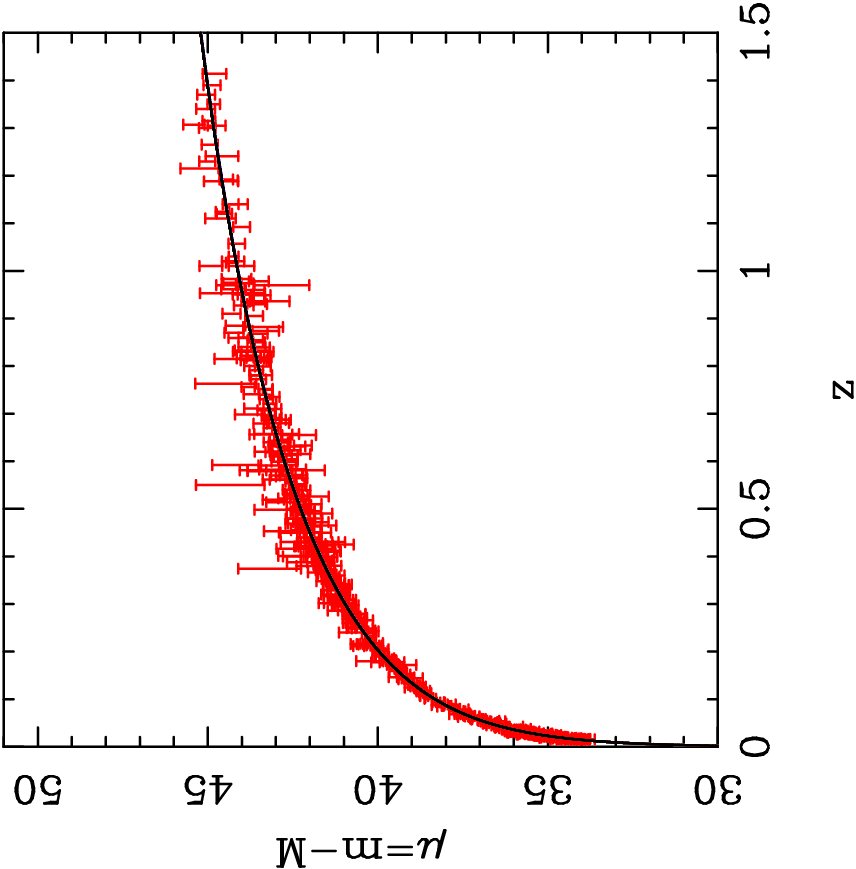}
        \caption{In this figure the acoustic parameter $A(z)$, the Hubble parameter $H(z)$ and 
	      the distance modulus $\mu(z)$ are shown in the plots from left to right respectively
	      as a function of redshift $z$ for inverse square potential~(\ref{eq:inverse_potential}).
	      The data points and error bars are taken from~\cite{Blake2011,Farooq2017,Suzuki2012}.
	      There are six very closely separated solid lines representing the model with inverse
	      square potential in each plot for the values of the parameter 
	      $\phi_0H_0=~2.0,~3.0,~4.0,~5.0,~6.0$ and $~7.0$.
	      The values of other parameters $\Omega_{m0}$ and $w_{\phi 0}$ are the corresponding
	      best fit values taken from each row of the table~\ref{tab:om_w_in}.
	      There is a good  agreement of the theoretical quantities with their observed values.}
        \label{fig:inv_alldata}
\end{figure}
\subsection{The Inverse Square Potential}
    \label{subsec:invrsepotential}
A potential which describes a tachyon scalar field model of dark energy is given as 
   \begin{equation}
	  V(\phi)=\frac{n}{4\pi G} \left(1-\frac{2}{3n}\right)^{1/2}\phi^{-2},
	  \label{eq:inverse_potential}
	\end{equation}
where the real number $n$ determines the amplitude of the potential. 
The inverse square potential leads to a cosmological evolution of the
form $a=t^n$ ~\cite{Paddy2002}.    
Cosmological dynamics of tachyon scalar field dark energy with this
potential have been studied in~\cite{BJP2003}, and the stability
analysis of this potential has been done
in~\cite{Copeland2005,Calcagni2006,Aguirregabiria2004}. 
The cosmological dynamics depends on the quantity
$\lambda = -M_n V^{-3/2}dV/d\phi$ which is a constant for this
potential.  
With the slow-rolling condition, this leads to a stable critical fixed
point for this potential which can generate a late time accelerated
expansion (with $n> 1$).  
This fixed point is an attractor which leads to $\Omega_\phi = 1$ and
the equation of state parameter $w_\phi=2/3n - 1$ asymptotically. 
There still remains the requirement of a tuning, which is needed for
a sufficient acceleration at the present
time~\cite{Copeland2005,Calcagni2006}.   

To numerically solve the cosmological equations, we transform the
above equations by introducing the following dimensionless variables 
 \begin{equation}
	 \begin{aligned}
	     y&=\frac{a(t)}{a(t_{in})}, \, \, \psi=\frac{\phi(t)}{\phi(t_{in})},\\
	     x&=H_{in}t,
	  \end{aligned}
	  \label{eq:dimensionless_variables}
    \end{equation}
here '$t_{in}$' represents the initial time.
The equations can then be written as
    \begin{equation}
	   \begin{aligned}
	      y^\prime &= y \left [\Omega_{m,in}y^{-3} + \Omega_{r,in}y^{-4} 
			+ \frac{\Omega_{\phi,in}\sqrt{-w_{\phi,in}}}
			{\psi^2\sqrt{1-\phi_{in}^2 H_{in}^2{\psi^\prime}^2}} \right ]^{1/2},\\ 
	      \psi^{\prime \prime} &= (1-\phi_{in}^2 H_{in}^2 {\psi^\prime}^2)
				      \left [\frac{2}{\phi_{in}^2H_{in}^2\psi}
				    - 3\psi^\prime \frac{y^\prime}{y} \right ],
	   \end{aligned}
	   \label{eq:dimensionless_eq_invrse}
    \end{equation}
The prime on superscript denotes derivative with respect to $x=H_{in}t$, and
different  $\Omega$'s are  dimensionless density parameters defined as
the ratio of the density of the relevant component and critical density
$\rho_{cr}=\frac{3H_0^2}{8\pi G}$. 
Here, we have assumed the universe to be spatially flat and hence
$\Omega_{total}=\Omega_{m,in} + \Omega_{r,in} + \Omega_{\phi,in} = 1$.  

\begin{figure}[t]
        \centering
        \includegraphics[width=0.32\textwidth,angle=-90]{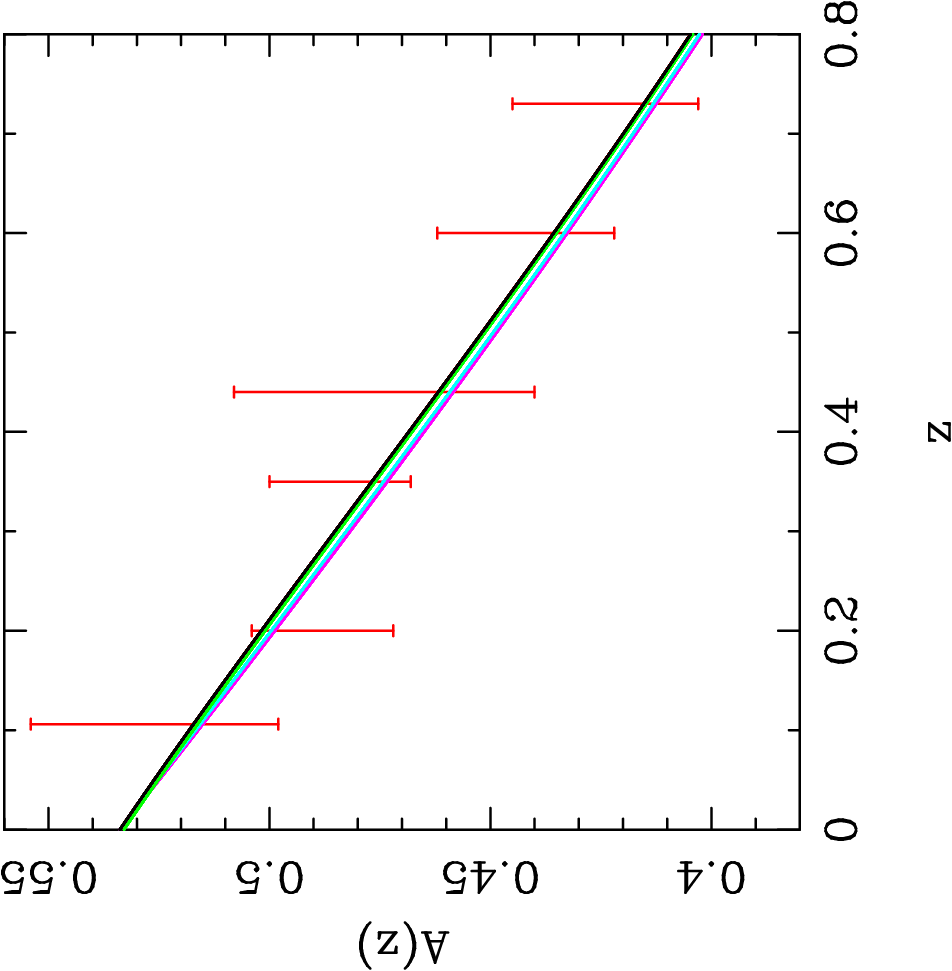}
        \includegraphics[width=0.32\textwidth,angle=-90]{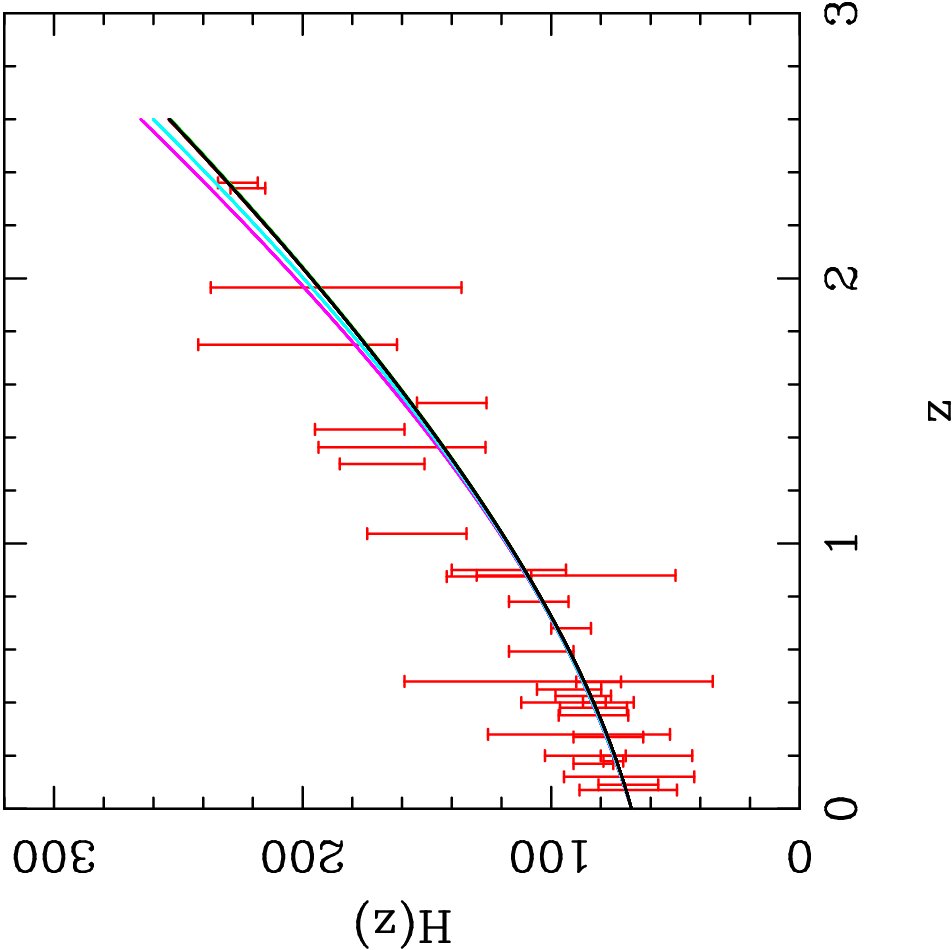}
        \includegraphics[width=0.32\textwidth,angle=-90]{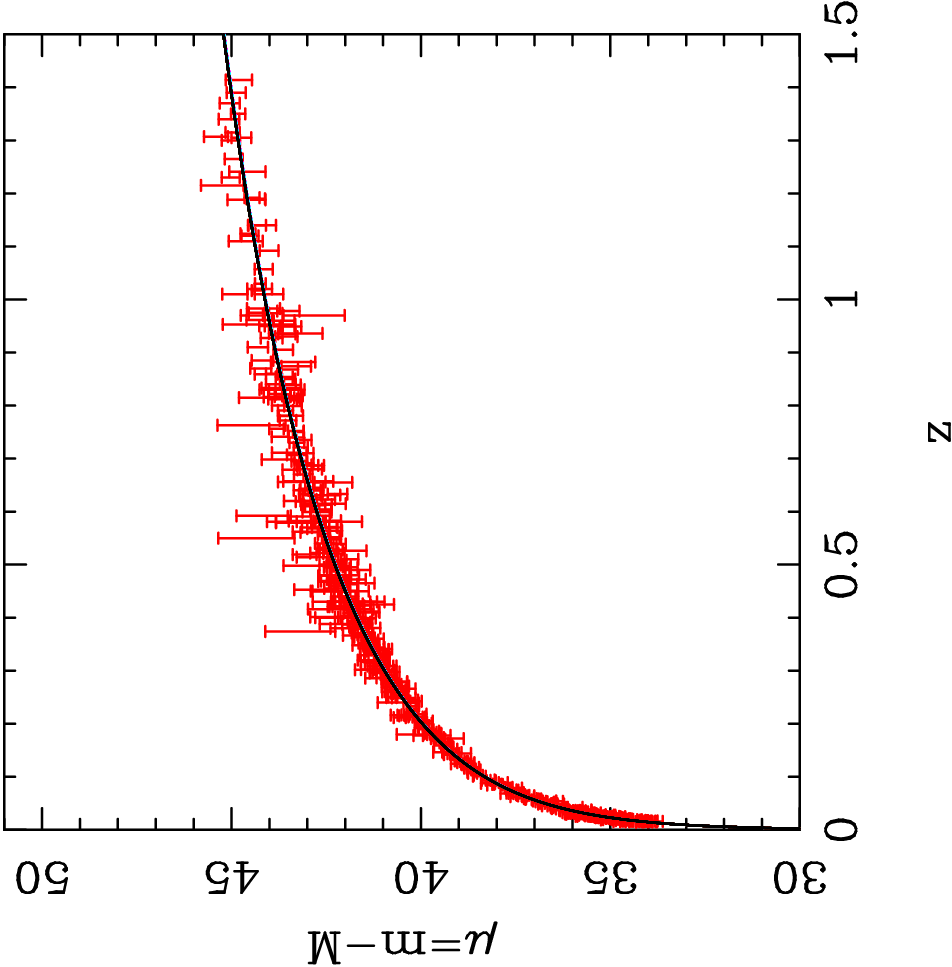}
        \caption{The figure shows plots of the acoustic parameter $A(z)$, the Hubble parameter $H(z)$ and
		 the distance modulus $\mu(z)$ as a function of redshift $z$ for the exponential potential
		 ~(\ref{eq:expo_potential}).
		 The data points and error bars are taken from~\cite{Blake2011,Farooq2017,Suzuki2012}.
		 There are six very closely separated theoretical solid lines representing tachyon dark 
		 energy model with exponential potential for $\phi_0H_0=~0.08,~0.09,~0.1,~0.3,~0.5$ 
		 and $~0.7$.
		 We have fixed the parameter $\phi_0/\phi_a=0.1$ and the values of other parameters
		 $\Omega_{m0}$ and $w_{\phi 0}$ are the corresponding best fit values taken from each row of
		 the table~\ref{tab:expo_omm_w}.} 
        \label{fig:expo_alldata}
\end{figure}
We integrate the equations numerically from the present time
($t_{in}=t_0$) to early times, and $\Omega_{m0}$, $\phi_0H_0$ and
$\dot{\phi_0}$ or $w_{\phi0}$ are the parameters which are varied. 
The amplitude of the potential can be constrained by using the relation
	\begin{equation}
	  \frac{2n}{3}\left(1-\frac{2}{3n}\right)^{1/2} = \Omega_{\phi 0}\phi_0^2H_0^2\sqrt{-w_{\phi 0}}.
	  \label{eq:amplitude_invrse}
	\end{equation}
To calculate the value of $n$ from above, the equation we need to solve
 is the polynomial equation 
	\begin{equation}
	  12n^3 -8n^2 -27q^2n = 0,
	  \label{eq:poly_ampl}
	\end{equation}
where $q=\Omega_{\phi 0}\phi_0^2H_0^2\sqrt{-w_{\phi 0}}$ is a
positive number.
The  solution of equation~(\ref{eq:poly_ampl}) for accelerated
expansion ($n>1$) is 
	  \begin{equation}
	    n = \frac{1}{3} + \frac{1}{6}\sqrt{4 + (9q)^2},  
	    \label{eq:value_n}
	  \end{equation}
with the condition that  $q > \frac{2\sqrt{3}}{9}$.	  
The value of the present day radiation density parameter $\Omega_{r0}$ is~\cite{WangAndWang2013}
\begin{equation}
    \Omega_{r0} = \frac{\Omega_{m0}}{1+z_{eq}},
    \label{eq:omr0}
\end{equation}
where $z_{eq}=2.5\times10^4\Omega_{m0} h^2(T_{cmb}/2.7K)^{-4}$, $T_{cmb} = 2.7255K $.\\ 
The initial conditions for the numerical solutions are
\begin{equation}
 y_0 = 1, \, \, \, \psi_0 = 1,
 \label{eq:initial_condi_1}
\end{equation}
and $\psi_0^\prime$ can be calculated using relation
    \begin{equation}
	    \psi^\prime = \frac{\dot{\phi}}{\phi_0 H_0} = \frac{\sqrt{1+w_\phi}}{\phi_0 H_0}
	    \label{eq:initial_condi_2}.
    \end{equation}
\subsection{The Exponential Potential}
    \label{subsec:expopotential}
The exponential potential for tachyon scalar field dark energy is given by
	\begin{equation}
	  V(\phi) = V_a \exp \left(-\phi/\phi_a \right),
	  \label{eq:expo_potential}
	\end{equation}
where amplitude $V_a$ and $\phi_a$ are the scalar field  parameters.    
Cosmological dynamics with this potential have also been studied
in~\cite{BJP2003}, and the stability analysis of this potential has
been done  in~\cite{Copeland2005,Calcagni2006,Aguirregabiria2004}.
For this potential, $\lambda \rightarrow \infty $ as $\phi \rightarrow
\infty$. 
This is a fixed point for which $\Omega_\phi \simeq 0$ and a 
dust like equation of state. 
Since $\lambda$ changes dynamically~\cite{Copeland2005}, the universe
goes to a temporary accelerated phase for $\lambda \lesssim 1$ and enters a
decelerated phase for $\lambda \gg 1$. 
In other words, the present day acceleration is temporary, and the universe
enters a phase of decelerated expansion once again. 
This evolution of the universe, therefore, avoids the future
event horizon problem. 

\begin{figure}[t]
       \begin{center}
        \includegraphics[width=0.32\textwidth,angle=-90]{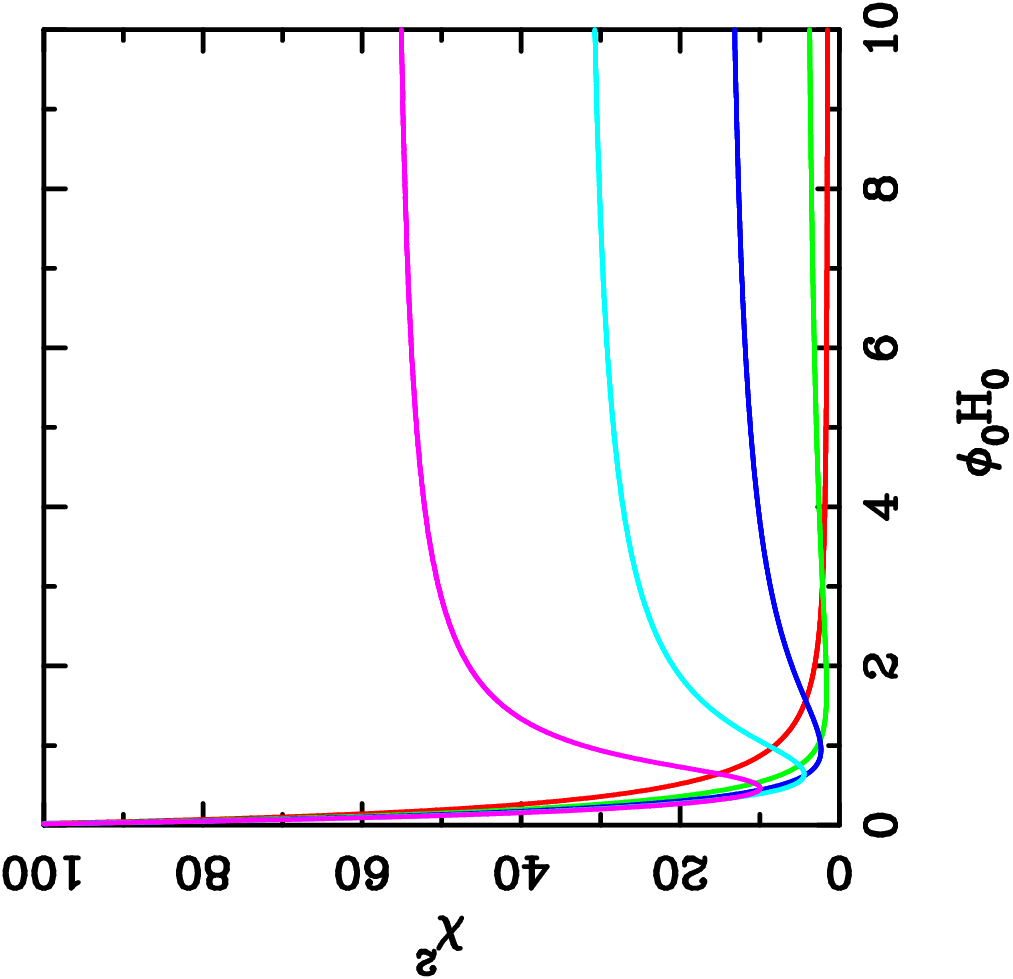}
         \includegraphics[width=0.32\textwidth,angle=-90]{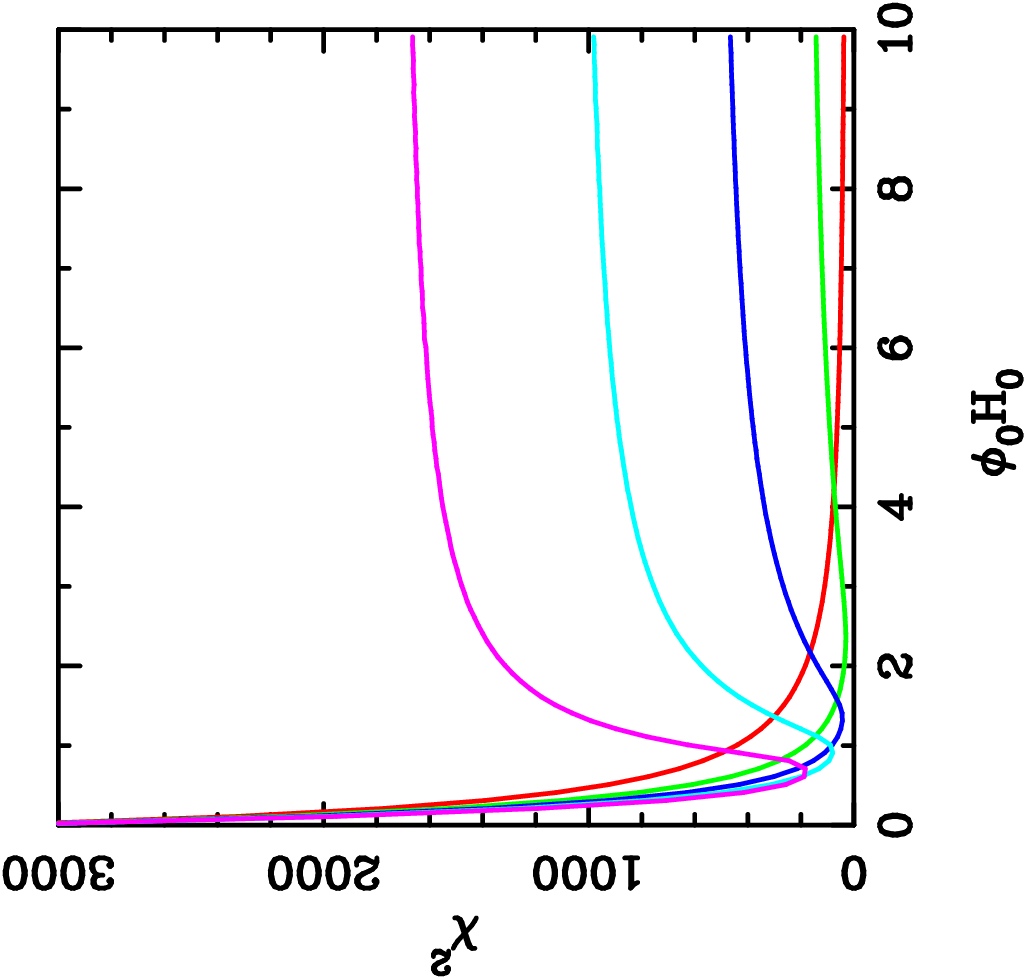}
         \includegraphics[width=0.32\textwidth,angle=-90]{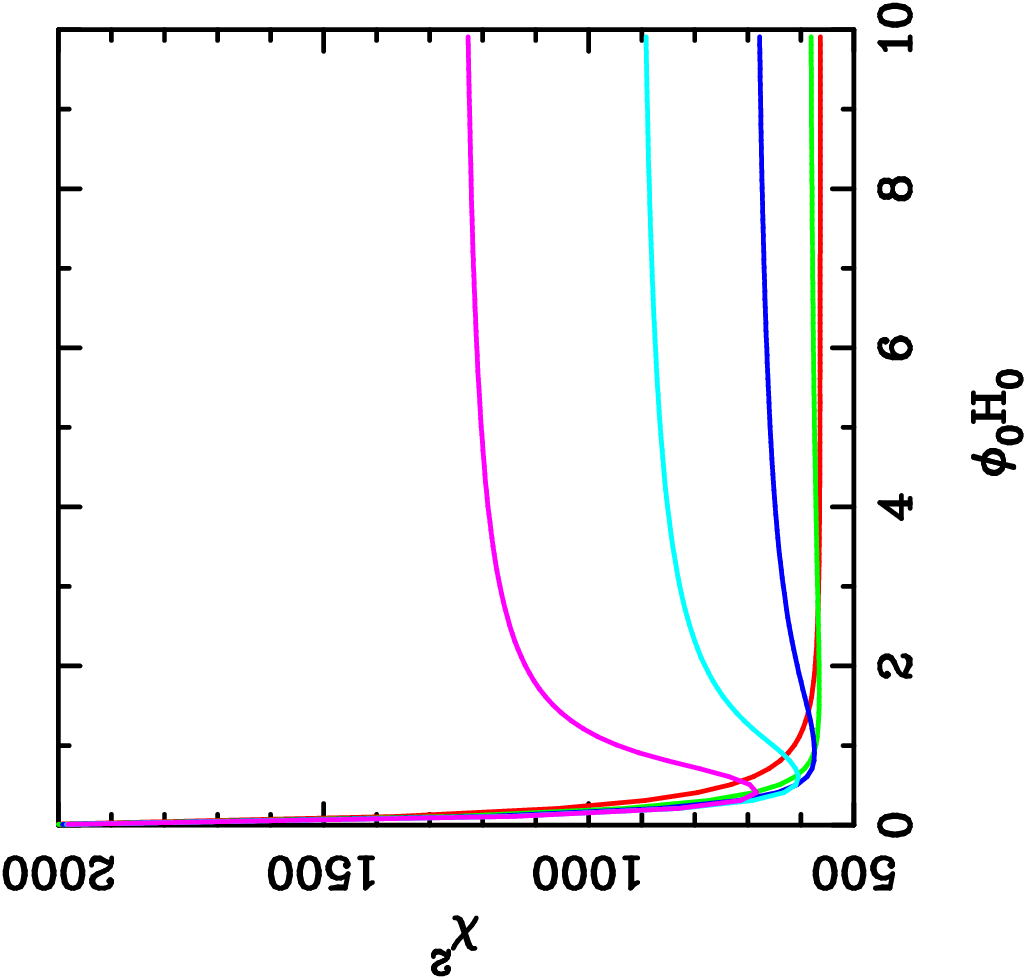}
         \caption{Going from left to right, the plots show $\chi^2$ as a function of
		  $\phi_0H_0$ for BAO, $H(z)$ and SN-Ia data respectively. Here have fixed
		  $\Omega_{m0}=0.285$ whereas the red, green, blue, sky-blue and pink lines
		  represent the value of $w_{\phi 0}$ to be $-1.0,~-0.95,~-0.90,~-0.85$ and $-0.80$
		  respectively.}
        \label{fig:chisqVSph0h0}
        \end{center}
    \end{figure}
Introducing the same dimensionless variables as introduced in the last
subsection, we can transform the required equations as
	 \begin{equation}
	    \begin{aligned}
	      y^\prime &= y \left [\Omega_{m,in}y^{-3} + \Omega_{r,in}y^{-4} 
			+ \frac{\Omega_{\phi,in}\sqrt{-w_{\phi,in}}e^{\frac{\phi_{in}}{\phi_a}(1-\psi)}}
			{\sqrt{1-\phi_{in}^2 H_{in}^2{\psi^\prime}^2}} \right ]^{1/2},\\ 
	      \psi^{\prime \prime} &= (1-\phi_{in}^2 H_{in}^2 {\psi^\prime}^2)
				      \left [\frac{\phi_{in}/\phi_a}{\phi_{in}^2H_{in}^2}
				      - 3\psi^\prime \frac{y^\prime}{y} \right ],\\
	    \end{aligned}
	    \label{eq:dimensionless_eq_expo}
	 \end{equation}
We, therefore, have three model parameters $\phi_0 H_0$ , $\phi_0/\phi_a$ and
$\dot{\phi_0}$ or $w_{\phi 0}$ to constrain. 
Apart from these parameters, there are cosmological parameters
$\Omega_{m0}$ and $H_0$. 
In this case, the amplitude of potential can be calculated by the relation
	    \begin{equation}
	      \frac{8\pi G}{3H_0^2}V_a = \Omega_{\phi 0}e^{\phi_0/\phi_a}\sqrt{-w_{\phi0}},
	      \label{eq:amplitude_expo}
	    \end{equation}
Structure of these equations suggest that $-1 \leq w_{\phi0} = \phi_0^2 H_0^2 {\psi_0^\prime}^2 -1 < 0$.
For this potential, we also use the same initial conditions given in
equation~(\ref{eq:initial_condi_1}) and (\ref{eq:initial_condi_2}).  
\section{Observations}
  \label{sec::observation}
  \begin{figure}[t]
        \centering
        \includegraphics[width=0.35\textwidth,angle=-90]{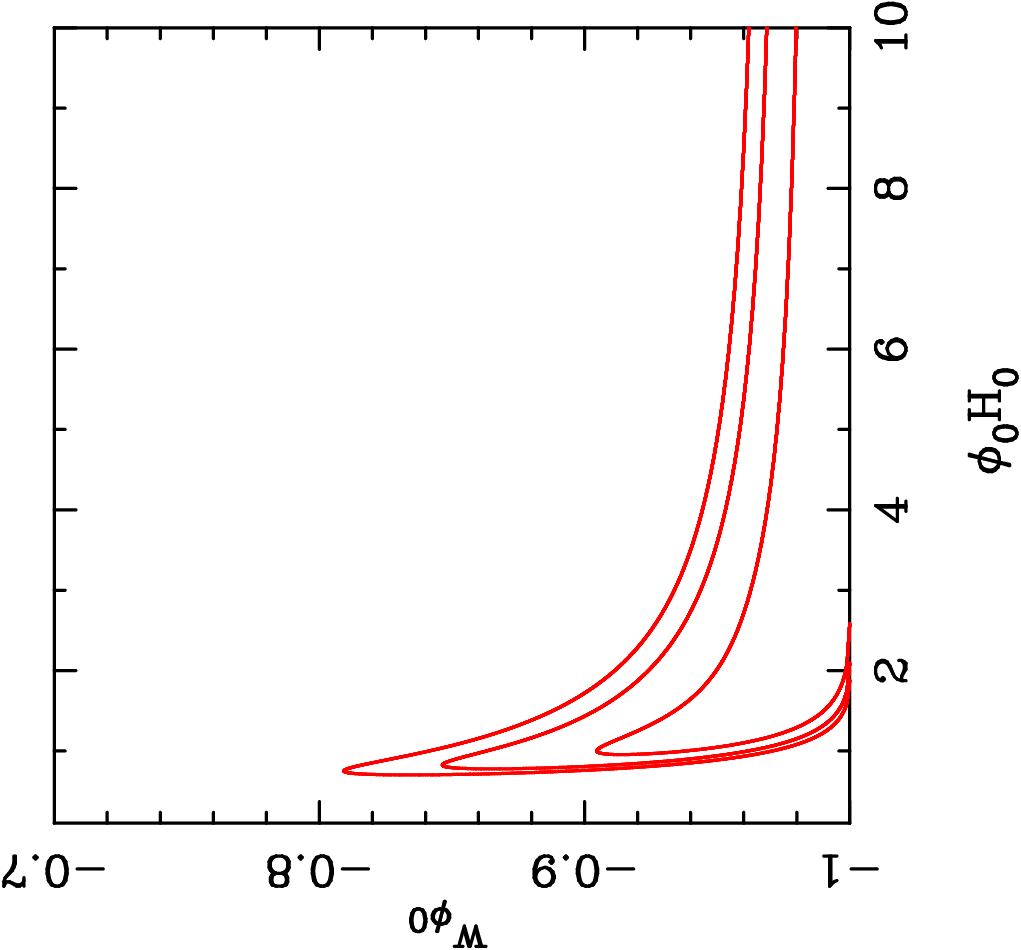}
        \includegraphics[width=0.35\textwidth,angle=-90]{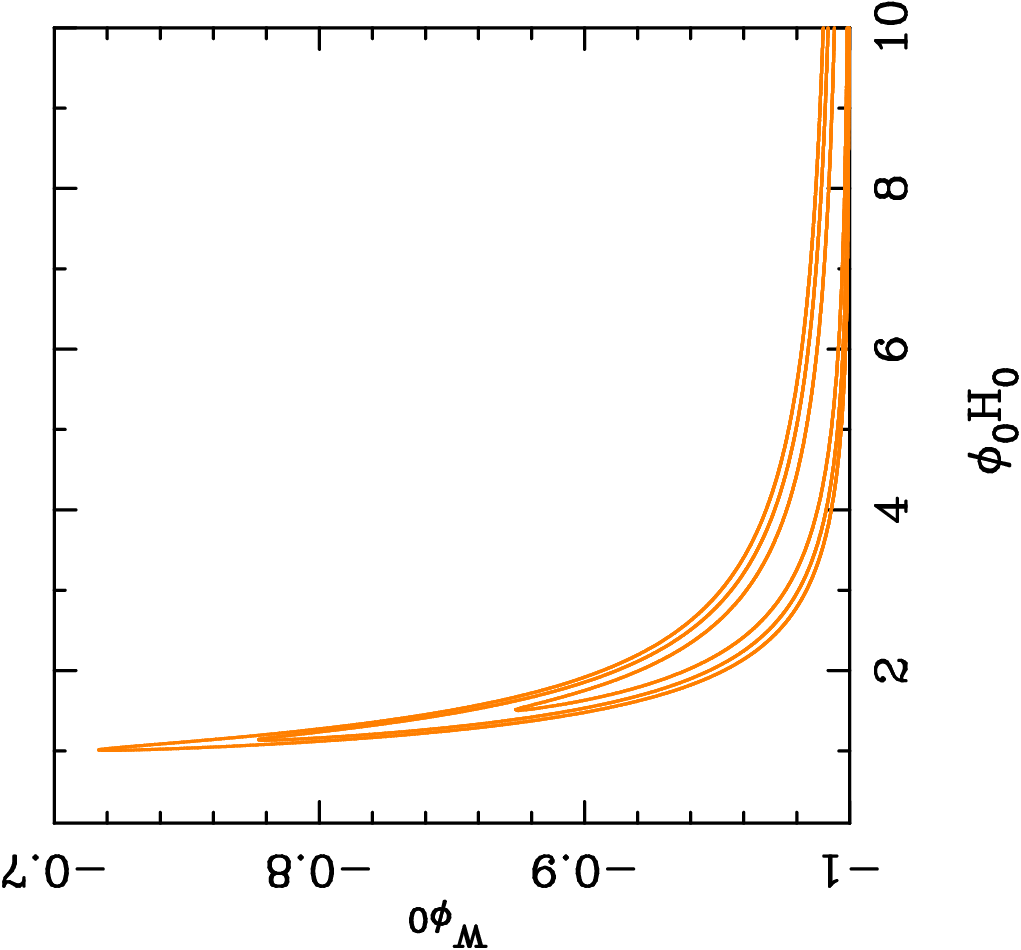}
        \includegraphics[width=0.35\textwidth,angle=-90]{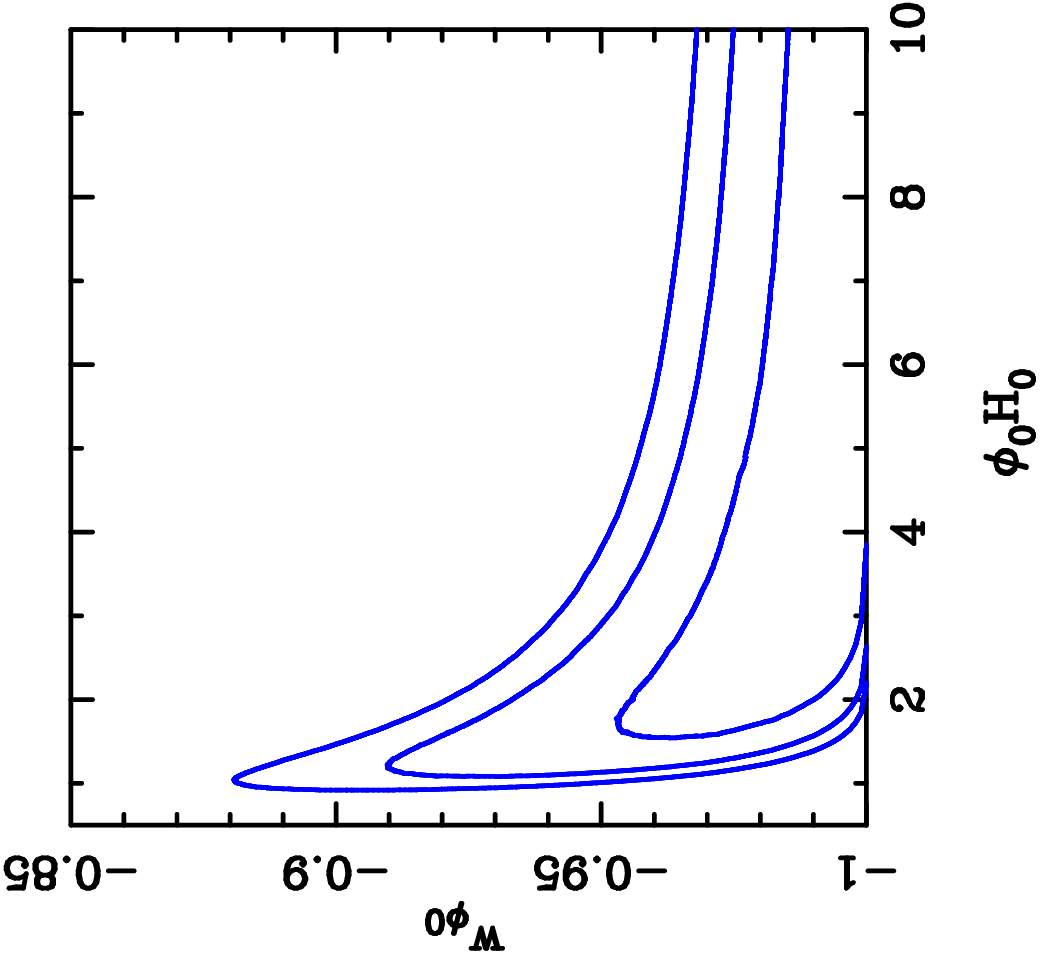}
        \caption{The figure shows $1\sigma$, $2\sigma$ and $3\sigma$ confidence contours between
		 $w_{\phi 0}$ and $\phi_0H_0$ for $\Omega_{m0}=0.285$ for BAO, $H(z)$ and SN-Ia
		 data in plots from left to right respectively.}
        \label{fig:invrse_wpvshph0}
    \end{figure}
\subsection{Baryon Acoustic Oscillation Data}
    \label{subsec:dataBAO}
Baryon Acoustic Oscillations
(BAO)~\cite{Seo2003,Percival2007,sca2013,Blake2012,Anderson2014,Veropalumbo2014,Delubac2015}
observations are the measurement of baryon oscillation feature in the
correlation function of large scale structure (LSS). 
This feature is the result of acoustic waves in the pre-recombination
baryon-photon plasma caused by opposing forces of gravity and
radiation. 
These acoustic waves left an imprint on the baryonic clustering in the
universe and gave rise to the baryon acoustic oscillation peaks. 
The characteristic angular scale of the acoustic peak is
$\theta_A = r_s(z_d)/D_V(z)$,
where $r_s$ is sound horizon at drag epoch $z_d$, which is
given by  
	  \begin{equation}
	    r_s(z_d)= \int^\infty_{z_d} \frac{c_s(z)}{H(z)}dz,
	    \label{eq:soundhorizon_def}
	  \end{equation}
and $D_V$ is effective distance ratio, and it can be calculated using the
angular diameter distance $D_A(z)$ as follows  
	    \begin{equation}
	      D_V(z)= \left [(1+z)^2 D_A(z)^2\frac{cz}{H(z)} \right]^{1/3}.
	      \label{eq:effective_dist_ratio}
	    \end{equation}
We use the BAO data from Baryon Oscillation Spectroscopic Survey
(BOSS) DR12 ~\cite{Alam2017} which provides 
6 data points (see table-7 of Alam et al.) at redshifts $z=~0.38,~0.51,~
0.61$ in terms of  $H(z)r_s(z_d)/r_{s,fid}$ and 
$D_M(z)r_{s,fid}/r_s(z_d)$ where $r_{s,fid} = 147.78$ Mpc and
$D_M(z)=(1+z)D_A(z)$ is the comoving angular diameter distance.  
The sound horizon $r_s(z_d)$ given by~\cite{Aubourg2015} 
\begin{equation}
    \label{eq:soundhorizon}
    r_s(z_d)=\frac{55.154\exp[-72.3(\omega_\nu + 0.0006)^2]}{\omega_b^{0.12807}
	     \omega_{cb}^{0.25351}}\rm{Mpc},
\end{equation}
where $\omega_\nu= \Omega_\nu h^2 = 0.0107(\sum m_\nu /
1.0 eV)$, $\omega_b = \Omega_bh^2 $ and $\omega_{cb}=\Omega_mh^2 -
\omega_\nu $. 
Symbols $\Omega_\nu$, $\Omega_b$ and $\Omega_m$ represent density parameters of
neutrinos, baryons and non-relativistic matter (baryonic matter + dark matter). 
We set mass of neutrinos $\sum m_\nu = 0.06$ eV and $\Omega_bh^2=0.02225$ with
$h=0.676$.  

We also use BAO data from LOWZ and CMASS at redshift $z=0.32$ 
 and $0.59$ as given in reference~\cite{Chuang-Hsun2017}. 
Here $r_{s,fid}=147.66 $ Mpc and the approximation for $r_s(z_d)$ is
the same as shown in equation~(\ref{eq:soundhorizon}).
We also use older BAO data from 6dFGS, SDSS DR7 and WiggleZ at redshifts
$z=0.106,~0.2,~0.35,~0.44,~0.6 $ and $~0.73$. 
These are listed in table-3 of~\cite{Blake2011} in term of the acoustic parameter. 
The acoustic parameter~\cite{Eisenstein2005} is 
	  \begin{equation}
	    A(z)= \frac{100D_A\sqrt{\Omega_m h^2}}{cz}.
	    \label{eq:accousticpara}
	  \end{equation}
Here $c$ is the speed of light in vacuum.
\begin{figure}[t]
    \begin{center}
         \includegraphics[width=0.33\textwidth,angle=-90]{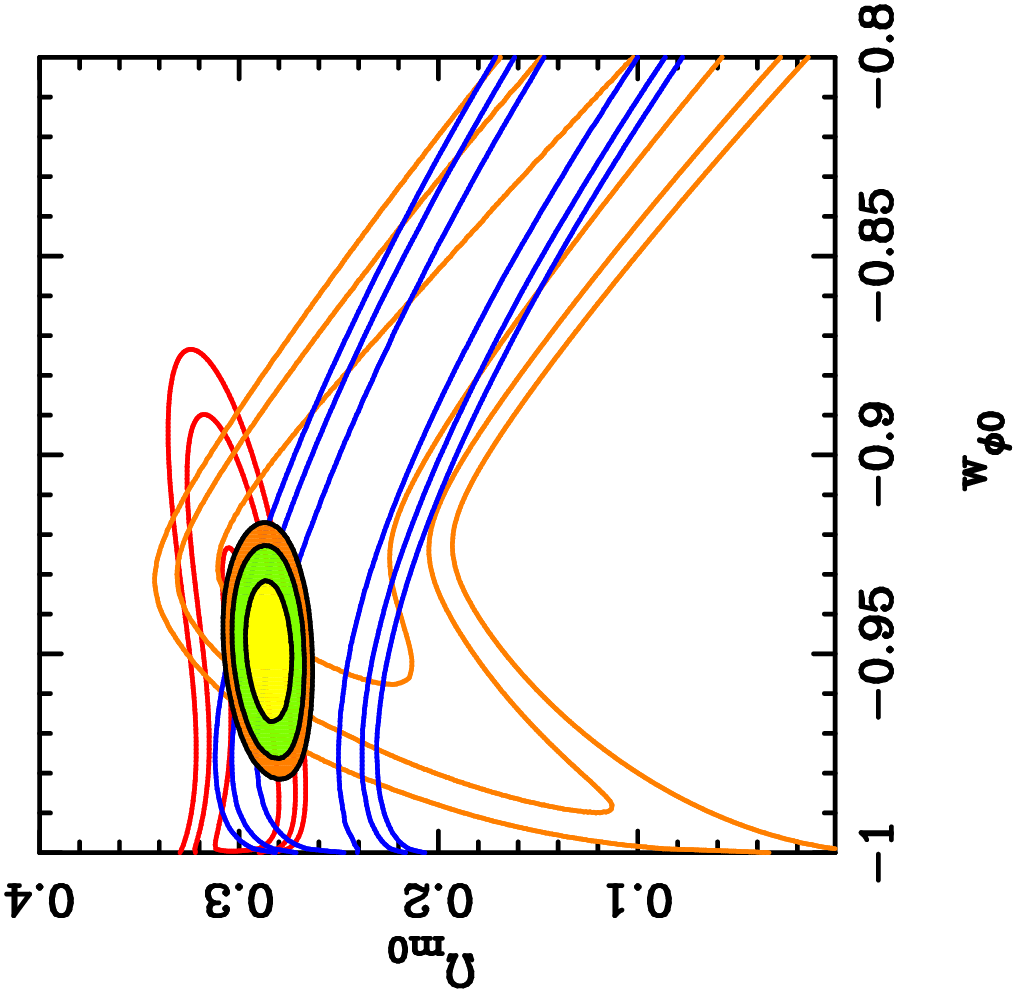}
         \includegraphics[width=0.33\textwidth,angle=-90]{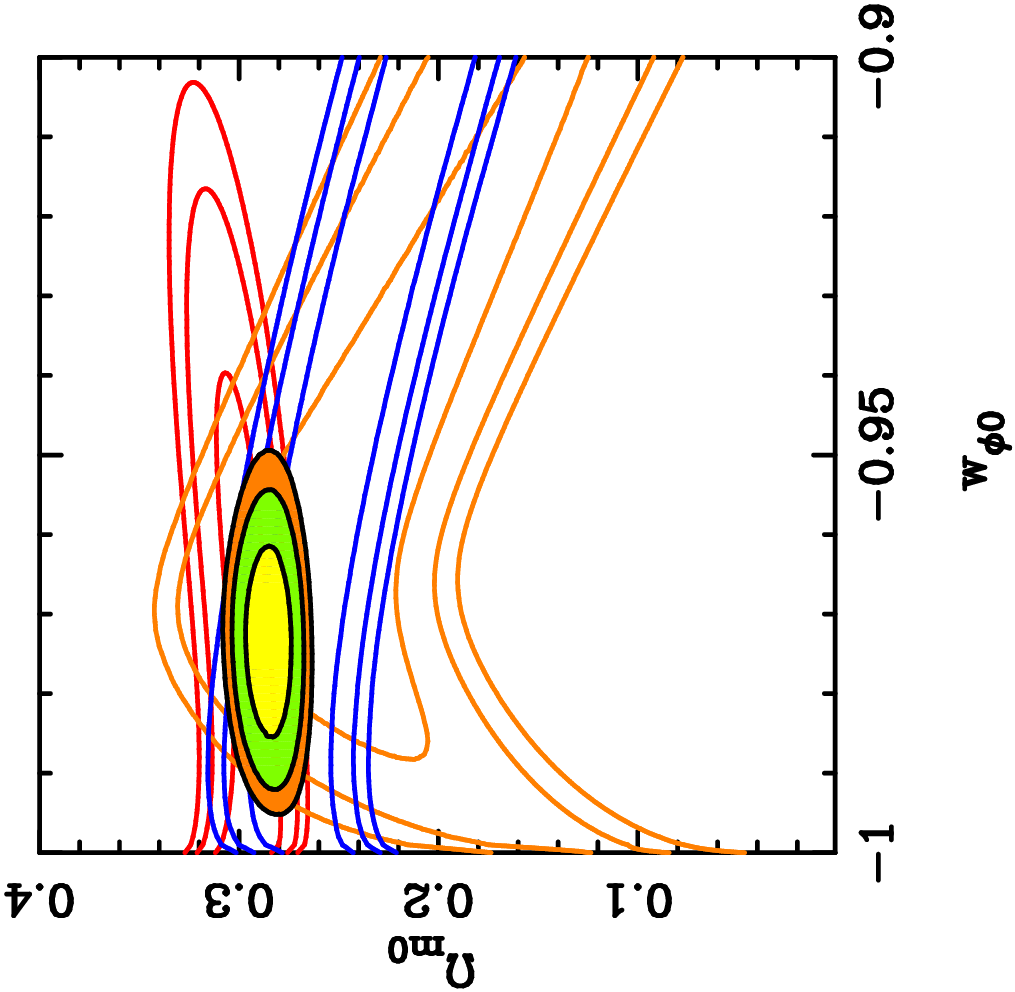}
         \includegraphics[width=0.33\textwidth,angle=-90]{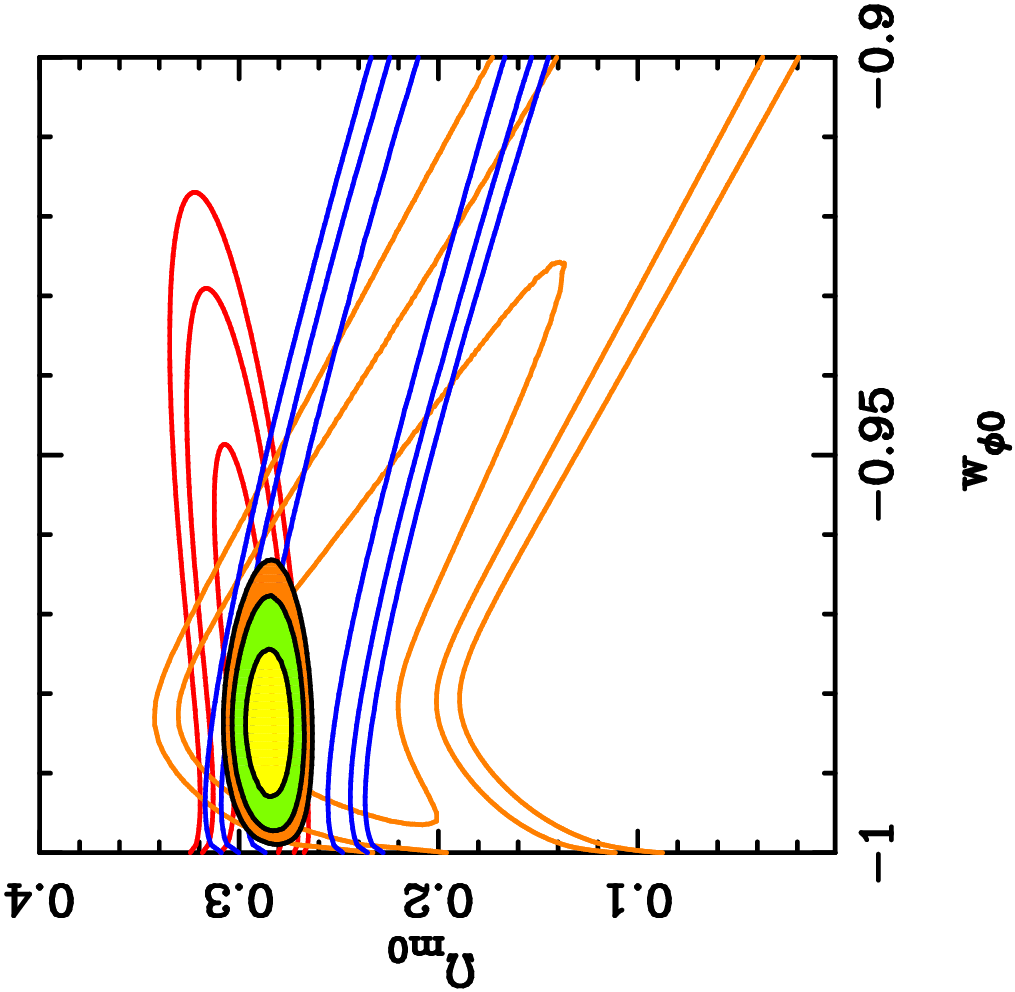}\\
         \includegraphics[width=0.33\textwidth,angle=-90]{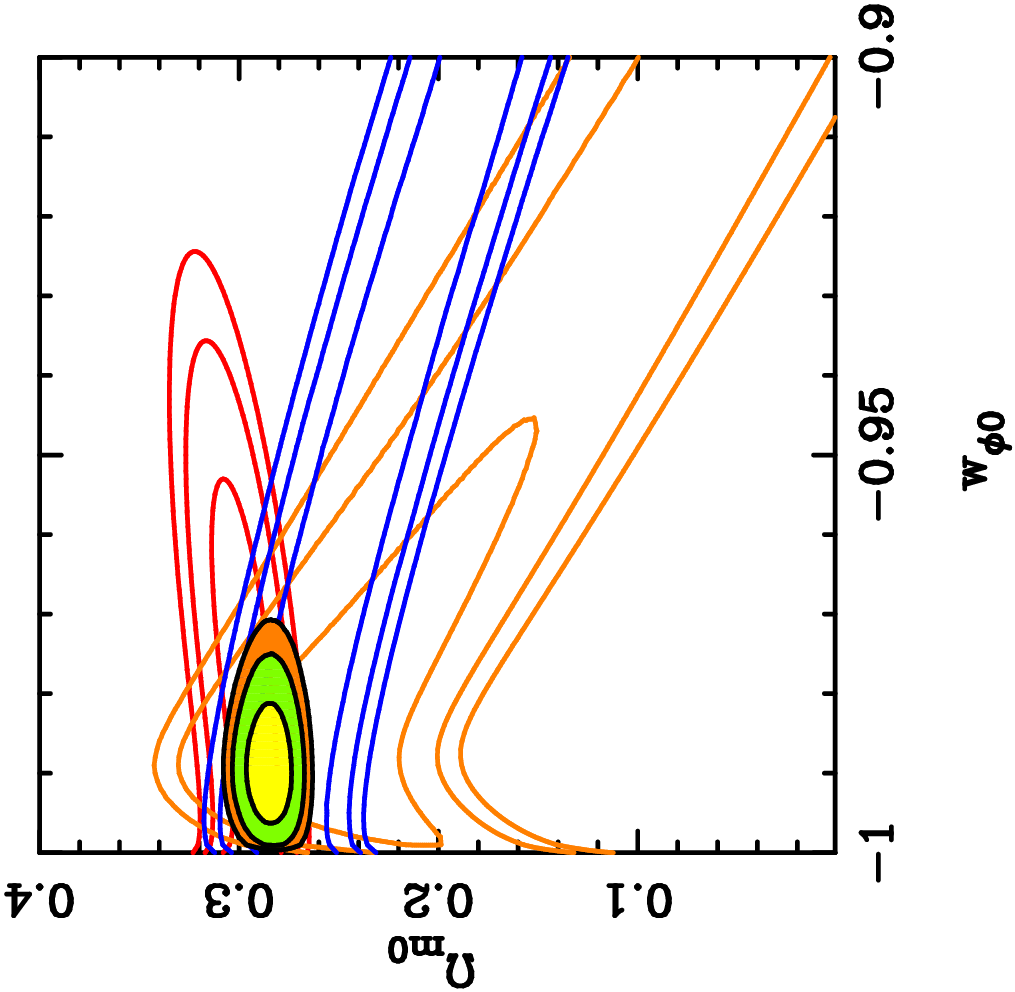}
         \includegraphics[width=0.33\textwidth,angle=-90]{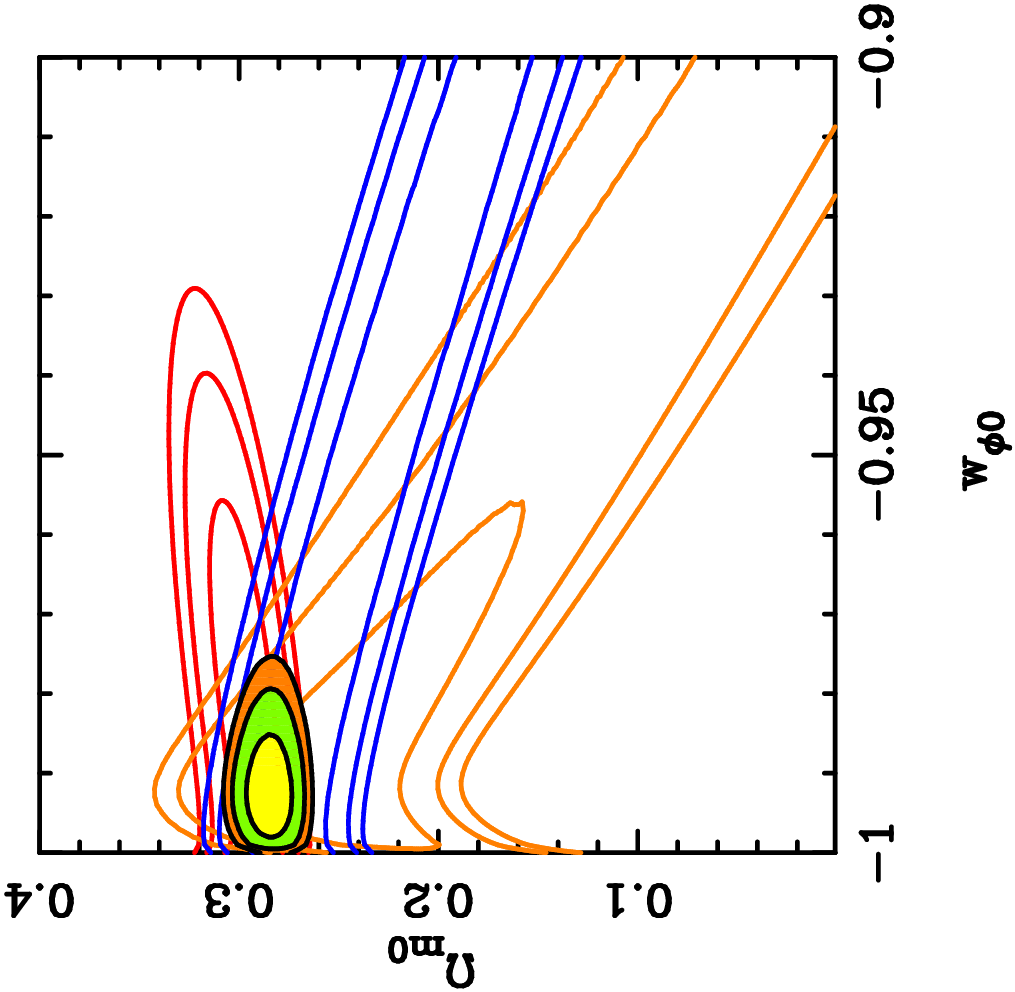}
         \includegraphics[width=0.33\textwidth,angle=-90]{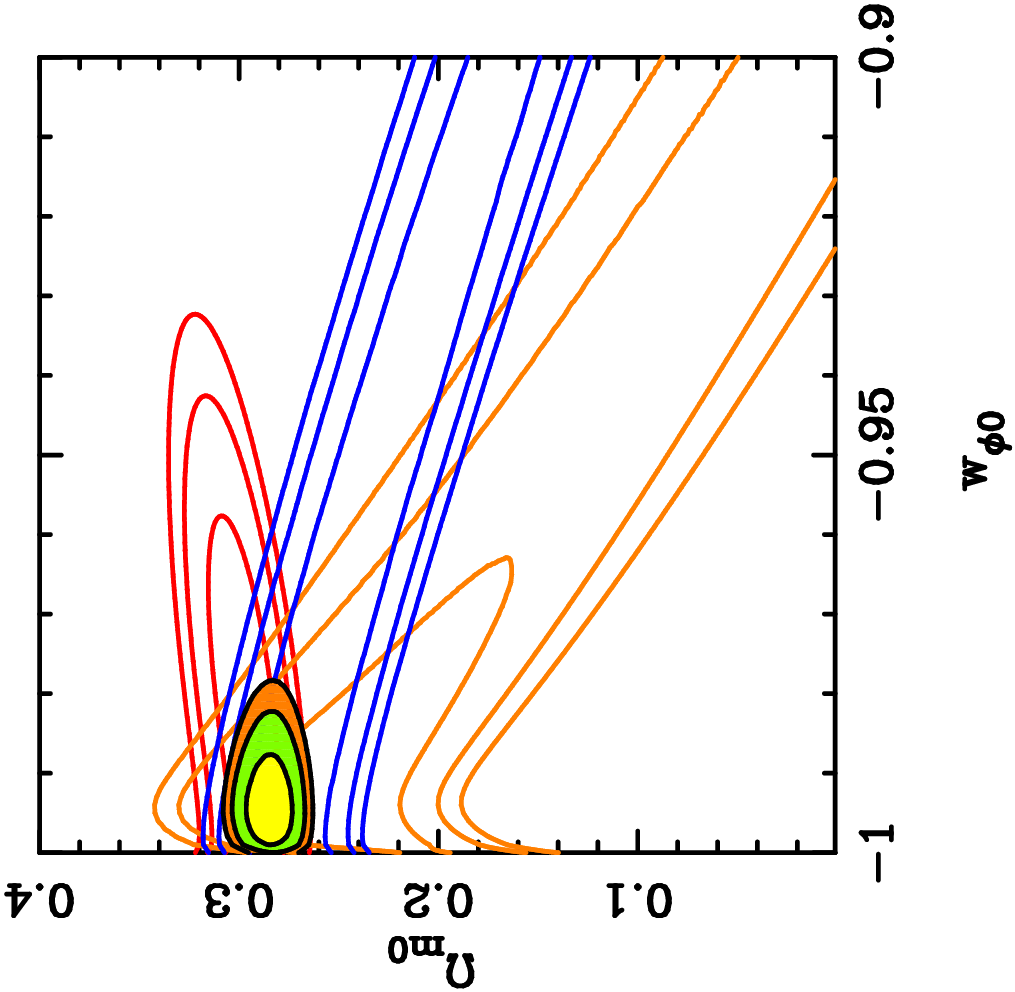}
         \caption{The plots show contours in the $\Omega_m - w_\phi$ plane, for constant
		  $\phi_0H_0$ for all the three and combined datasets for the inverse square potential.
		  Contours in red, orange and blue  are for BAO, $H(z)$ and SN-Ia data respectively.
		  Black contours filled with colours represent combined constraints.
		  The value of parameter $\phi_0H_0=~2.0,~3.0,~4.0$ for plots in row-1 
		  and $5.0,~6.0,~7.0$ for plots in row-2 from left to right respectively.}
        \label{fig:incntromwp}
    \end{center}
  \end{figure}
  \begin{table}[t]
        \centering
       \begin{tabular}{|c|c|c|c|c|}
	  \hline
	  &  &  & &  \\
	  $\phi_0H_0$ & $\chi^2_{min}$ & $\Omega_{m0}$ & $w_{\phi 0}$ &  $n$ \\
	  &  &  & &  \\
	  \hline
	  &  &  & &  \\
	  2.0 & 596.145 & $0.285^{+0.023}_{-0.022}$ & $-0.950^{+0.033}_{-0.031}$ & [4.323,4.726]  \\
          &  &  & &  \\
          \hline
          &  &  & &  \\
          3.0 &  592.045 & $0.285^{+0.023}_{-0.021}$  & $-0.973^{+0.023}_{-0.022}$ & [9.445,10.250] \\
          &  &  & &  \\
         \hline
	 &  &  & &  \\
	 4.0 & 590.944  & $0.284^{+0.024}_{-0.021}$ & $-0.984^{+0.021}_{-0.015}$ & [16.635,18.016] \\
         &  &  & &  \\
         \hline
          &  &  & &  \\
	 5.0& 590.515 & $0.285^{+0.023}_{-0.022}$  & $-0.990^{+0.019}_{-0.009}$  & [25.906,27.959] \\
         &  &  & &  \\
         \hline
          &  &  & &  \\
	 6.0 & 590.335 & $0.285^{+0.023}_{-0.022}$ & $-0.993^{+0.017}_{-0.007}$ & [37.252,40.133] \\
         &  &  & &  \\
            \hline
         &  &  & &  \\
         7.0  & 590.285 & $0.285^{+0.023}_{-0.022}$ & $-0.995^{+0.016}_{-0.005}$ & [50.659,54.504] \\
         &  &  & &  \\
         \hline
        \end{tabular}
        \caption{The table lists the best fit values of $\Omega_{m0}$ and $w_{\phi0}$ along with their
		 $3\sigma$ confidence range for different values of $\phi_0H_0$ for 
		  inverse square potent for combined data (BAO + H(z) + SN-Ia). In the second
		  column minimum value of corresponding $\chi^2_{min}$ have been shown. 
		  In the last column, we have shown the $3\sigma$ allowed range of '$n$', 
		  calculated from equation~(\ref{eq:value_n}) considering $3\sigma$ confidence
		  range of $\Omega_{m0}$ and $w_{\phi0}$.}
        \label{tab:om_w_in}
    \end{table} 
\subsection{Hubble parameter Data}
    \label{subsec:dataHz}
Hubble parameter data
set~\cite{Farooq2017,Farooq2013a,Farooq2013b,Samushia2006,Stern2010,Moresco2012,Chuang2013,Chen2011}
consists of values of Hubble parameters $H(z)$ at different redshifts
and associated errors in the corresponding measurement.
We use $H(z)$ dataset compiled and listed in table-1
of~\cite{Farooq2017}.
The table contains values of the Hubble parameters $H(z)$ at $38$ different redshifts with
associated errors in measurement and corresponding references up to redshift $z=2.3$.
Out of $38$ we use only $32$ points as we do not consider three data points taken from 
Alam et. al.(2016) at redshifts $z=~0.38,~0.51,~0.61$ and three data points taken from
Black et. al.(2012) at redshifts
$z=~0.44,~0.6,~0.73$ as we include these data points in our BAO dataset.
Hubble parameter can be computed from the Friedmann equation and is given
by 
    \begin{equation}
        \label{Hzvalue}
        H(z)= H_0 \left[\Omega_{m0} (1+z)^3 + \Omega_{r0}(1+z)^4 + \Omega_\phi \right]^{1/2},
    \end{equation}
where $H_0$ is the present value of the Hubble parameter.

\subsection{Supernova Type Ia Data}
    \label{subsec:dataSN}
\begin{figure}[t]
    \centering
    \includegraphics[width=0.35\textwidth,angle=-90]{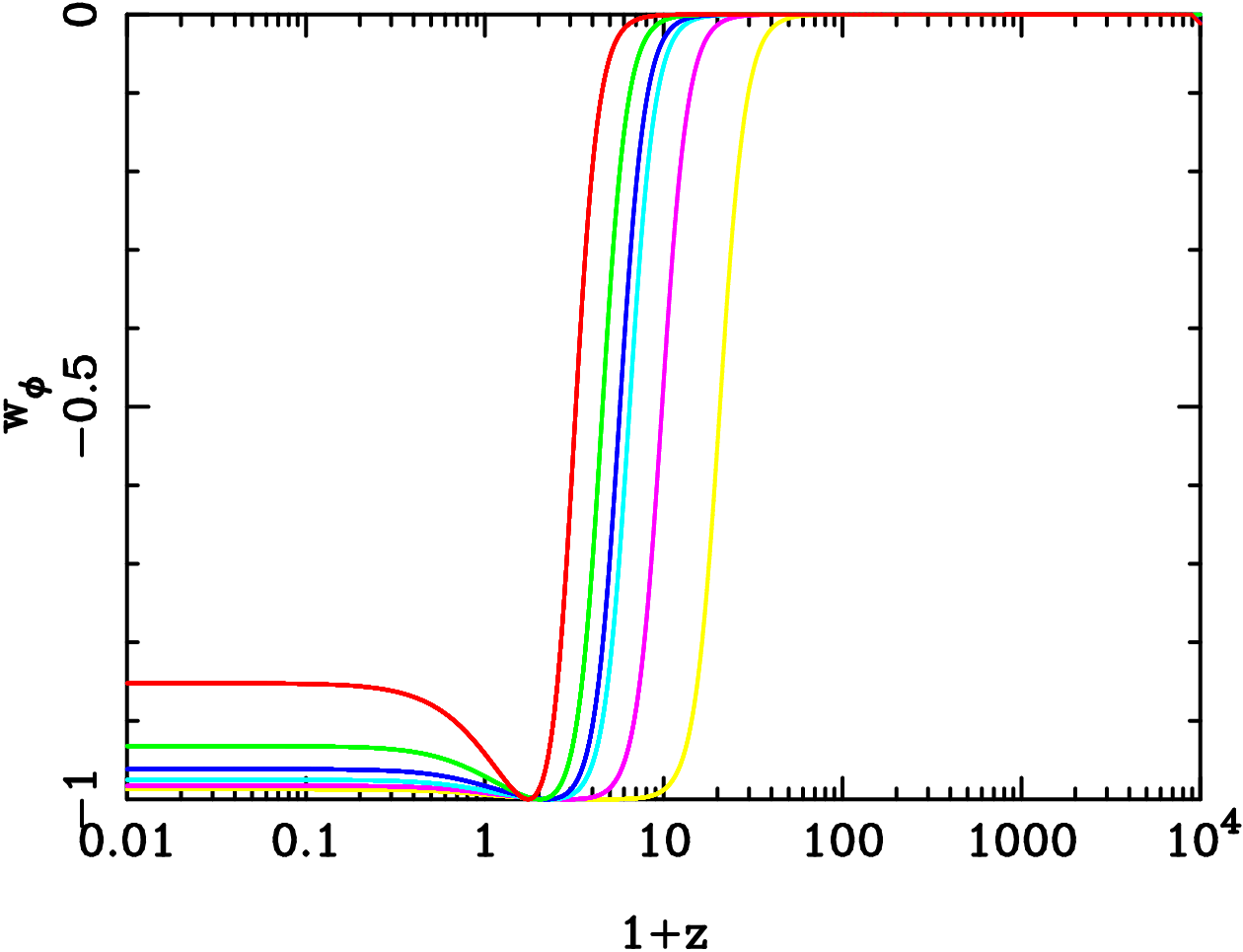}
    \includegraphics[width=0.35\textwidth,angle=-90]{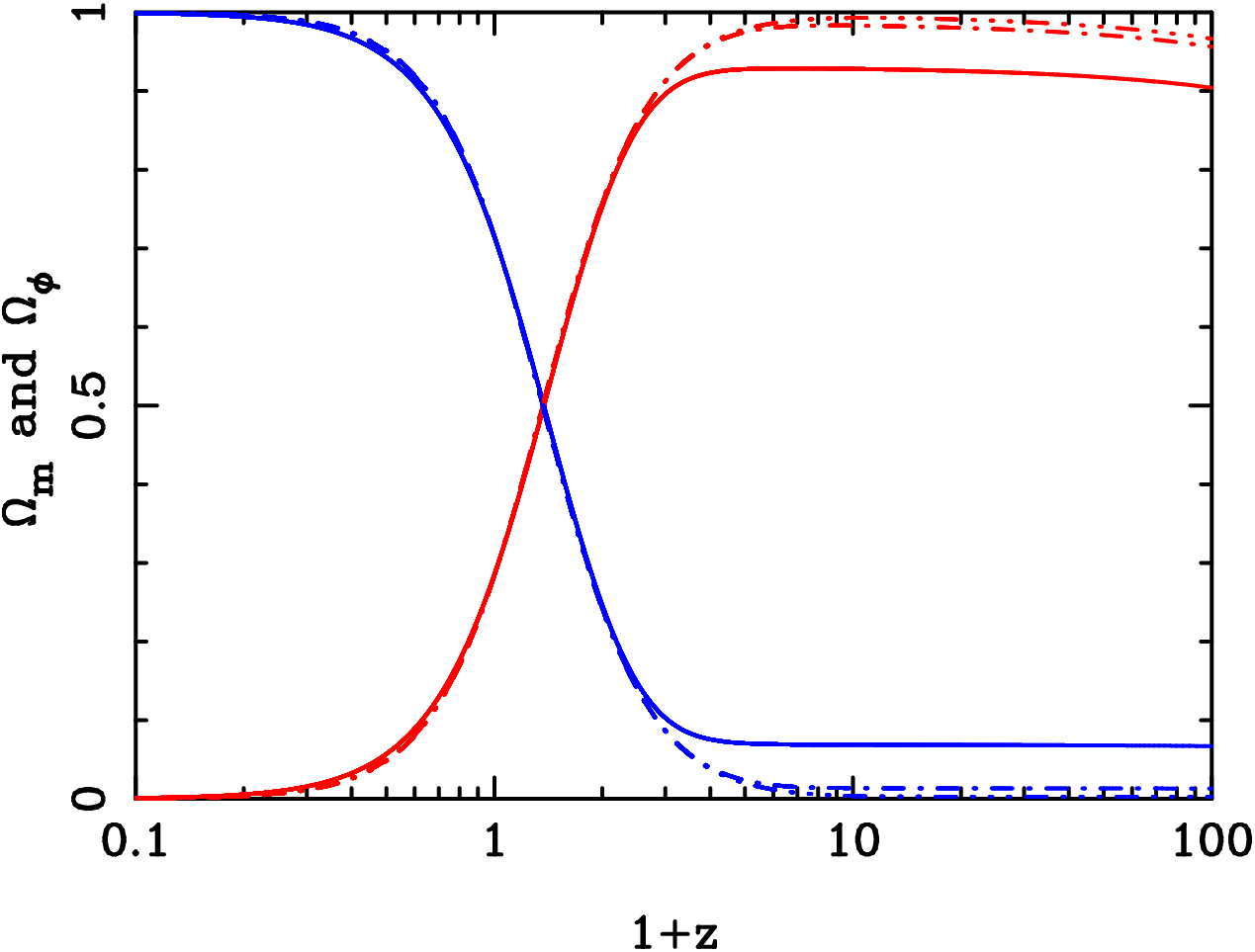}
    \caption{The plot on the left shows the evolution of equation of state $w_\phi$ with redshift
	     for inverse square potential. Red, green, blue, sky-blue, yellow and pink 
	     colours represent value of $\phi_0H_0=~2.0,~3.0,~4.0,~5.0,~6.0$ and $~7.0$.
	     The plot on the right shows the evolution of $\Omega_m $(red curves)
	     and $\Omega_\phi $(blue curves) with redshift.
	     The solid, dash-dot and dashed-dot-dot-dot lines represent the value of
	     $\phi_0H_0=~2.0,~4.0 $ and $6.0$ respectively.
	     The value of parameter $w_{\phi 0}$ and $\Omega_{m0}$ are the best fit values
	     taken from table~\ref{tab:om_w_in} for the corresponding value of $\phi_0H_0$.}
    \label{fig:evolution_invrse}
\end{figure}
The third dataset we use for our analysis is observations of supernovae type
Ia~\cite{Suzuki2012,Perlmutter1997,Perlmutter1999,Riess1998,Garnavich1998,Tonry2003,
Barris2004,Goobar2000,Gonzalez2012,Astier2006},
which is supernova explosion of a white dwarf star accreting mass from
its binary companion and hence crosses the Chandrasekhar limit.
When a white dwarf star crosses the Chandrasekhar mass limit of $1.4 M_{\odot}$, 
it explodes: this is a Supernova explosion of Type-Ia.
Luminosity distance of Supernova-Ia which occurred at redshift $z$ is given
by the relation
    \begin{equation}
        d_L(z)=\frac{c}{H_0}(1+z)\int_0^z \frac{1}{E(z)} dz.
        \label{eq:luminosity_dist}
    \end{equation}
    where function $E(z)=H(z)/H_0$.
In SN-Ia data, we have distance moduli of 580 supernovae up to
redshifts $z=1.414$ along with their associated observational
error~\cite{Suzuki2012}. 
The theoretical values of distance modulus can be calculated using
luminosity distance as 
    \begin{equation}
        \label{distancemodulus}
        \mu = 5 log(d_L) - 5,
    \end{equation}
here $d_L$ is in the unit of $10~pc$ and $\mu=m-M$ is the distance modulus, $m$ and $M$ are 
the apparent and absolute magnitude respectively of the supernova.
\section{Observational Constraints}
  \label{sec::results}
We do the standard $\chi^2$ analysis to constrain parameters for
the tachyon dark energy.
Value of $\chi^2_{BAO}$ for  the Baryon Acoustic Oscillation data is
the sum of  $\chi^2$ over all redshifts given in
subsection~\ref{subsec:dataBAO}. 
We calculated $\chi^2$ for DR12 data using the expression given by
\begin{equation}
    \chi^2 = \sum^N_{i,j=1}[X_{th,i}-X_{obs,i}]C^{-1}_{i,j}[X_{th,j}-X_{obs,j}],
\end{equation}
where $N$ is the number of data points in BAO dataset, {\bf $X_{th}$} is a
vector of the theoretical value of corresponding observable and {\bf
  $X_{obs}$} is a vector of the observational values.
We employ the covariance matrix $C_{ij}$ taken from the online files
of Alem et al. (2017) and Chi-Hsun et al. (2017).
Value of $\chi^2$ for older BAO data (BAO data from 6dFGS, SDSS DR7 and
WiggleZ), H(z) data and SN-Ia data is calculated using
    \begin{equation}
        \chi^2_{oder BAO/Hz/SN} = \sum^N_{i=1}\left (\frac{O_D(z_i)-O_M(z_i,\bf{p})}{\sigma_i}\right)^2, 
    \end{equation}
Here $O_D(z_i)$ is the theoretical value of the observable at redshift
$z_i$, and $O_M(z_i,\bf{p})$ is its value for model at redshift
$z_i$ with the set of parameters $\bf{p}$.
The quantity $\sigma_i$ is the error in the measurement of the observable $O_D(z_i)$.
Here observable $'O'$ is the acoustic parameter $A(z)$ for the older BAO data,
Hubble parameter for the H(z) data and distance modulus $\mu(z)$ for
the SN-Ia data.
We then find the maximum likelihood ($e^{-\chi^2_{tot}}$) of the parameter
space by minimizing   
$\chi^2_{tot} = \chi^2_{BAO} + \chi^2_{Hz} + \chi^2_{SN}$.

\subsection{Constraints on Inverse Square Potential}
    \label{subsec:constraint_on_iverse_pot}
As mentioned in section~\ref{subsec:invrsepotential}, we constrain
three parameters, $\Omega_{m0}$, $w_{\phi 0}$ and $\phi_0H_0$ for the
potential. 
Since only the square of the quantity $\phi_0H_0$ appears in the
equations, we need to consider only one of the two, positive or
negative branches.   
There is a degeneracy between parameters $w_{\phi 0}$ and
$\phi_0H_0$; these parameters are correlated.  

In figure~\ref{fig:inv_alldata}, we plot the acoustic parameter $A(z)$
obtained from the BAO data from 6dFGS, SDSS DR7 and WiggleZ~\cite{Blake2011}. 
In the plot on middle and on the right we have shown the Hubble parameter $H(z)$ 
and the distance modulus $\mu(z)$ as a function of $z$ respectively.
Data points and error bars are taken from~\cite{Farooq2017,Suzuki2012}.
There are six (overlapping) theoretical solid lines in each of these plots representing inverse 
square potential~(\ref{eq:inverse_potential}). 
To draw these theoretical lines we have taken the best fit value of parameters 
$\phi_0H_0$, $\Omega_{m0}$ and $w_{\phi0}$ from each row of table~\ref{tab:om_w_in}.
We can see that there is a good agreement of the theoretical curves with data. 

The values of $\chi^2$ vs $\phi_0H_0$ for
$\Omega_{m0}= 0.285 $ are plotted in   figure~\ref{fig:chisqVSph0h0}.   
The five different colours represent different values of $w_{\phi 0}$ from
$-1.0$ to $-0.80$ in the steps of $0.05$.  
We can see that if $w_{\phi 0}$ is close to $-1.0$ (red and green
curves) all larger value of $\phi_0H_0$ are allowed.  
If we fix $w_{\phi 0}$ to a value away form $-1.0$, we can get a
minimum in $\chi^2$ curves and fixing this parameter is equivalent to
fixing $\dot{\phi}_0$ as $w_\phi = \dot{\phi}^2 -1$. 
Since we are interested in constraining $\Omega_{m0}$ and $w_{\phi0}$,
we choose to fix $\phi_0H_0$.  
Degeneracy between these parameters can also be seen in
figure~\ref{fig:invrse_wpvshph0} where we have shown $1\sigma$,
$2\sigma$ and $3\sigma$ contours in the $w_{\phi 0}$ - $\phi_0H_0$
plane for the three datasets.
After marginalizing over $\Omega_{m0}$, we find that $\phi_0H_0 \ge 0.775$ 
at $3\sigma$ confidence level using combined data.
It can be clearly seen that there is a bound on the lower value of
$\phi_0H_0$ but not on its upper value and hence we
can not marginalize over $\phi_0H_0$.
We, therefore, have constrained the parameter space of
$\Omega_{m0}-w_{\phi 0}$ and shown its variation with $\phi_0H_0$
 in figure~\ref{fig:incntromwp}. 

For each of these contours, we have fixed the value of $\phi_0H_0$.
The most stringent constraints come from the BAO data, and combined
constraints limit the parameter space to a very small range. 
Value of $\Omega_{m0}$ is well constrained by combined dataset at
$0.285^{+0.023}_{-0.022}$ with $3\sigma$ confidence, and this remains
almost same with variation in parameter $\phi_0H_0$.
However, the constraint on $w_{\phi_0}$ depends on the value of $\phi_0H_0$.
As we increase the value of $\phi_0H_0$, all the three datasets prefer a
value of $w_{\phi 0}$ close to $-1$.  
In table~\ref{tab:om_w_in} we have shown, the minimum value of 
$\chi^2 $ for a fixed value of $\phi_0H_0$ and the best fit
value of parameters $\Omega_{m0}$ and $w_{\phi 0}$ with the 
$3\sigma$ confidence limit for combined data.
We started with $\phi_0H_0=2.0 $ and increased its value in unit step.
Here we can see that minimum value of $\chi^2$ saturates for a larger
value of $\phi_0H_0$, and so does the parameter $\Omega_{m0}$. 
In this background cosmological model, we can tune the parameter $\phi_0H_0$
to be very close to $w_\phi=-1.0$. 
In the strict sense, it is not possible to constrain $\phi_0H_0$ using
these background data. 
A large range of values of $\phi_0H_0$ are acceptable as the background
data only put a lower bound on its value.
In the last column of table~\ref{tab:om_w_in}, we have shown the
$3\sigma$ allowed range of '$n$' computed from the
equation~(\ref{eq:value_n}) considering the $3\sigma$ 
confidence range of $\Omega_{m0}$ and $w_{\phi0}$.
From equation~(\ref{eq:amplitude_invrse}), it is clear that the amplitude
of the tachyon potential and constant '$q$' are proportional to the value of
$\phi^2_0H^2_0$, as can be seen in the equation~(\ref{eq:amplitude_invrse}).
This is the reason, the allowed value of '$n$' also increases with it.
Since the universe expands like $a \propto t^n$ for a given '$n$', for a
larger value of $\phi_0H_0$ the accelerated expansion is faster in dark
energy dominated era. 
We find that for this model, the transition redshift is between $0.61
\lesssim z_{acc}\lesssim 0.80$. 
\begin{figure}[t]
    \centering
    \includegraphics[width=0.33\textwidth,angle=-90]{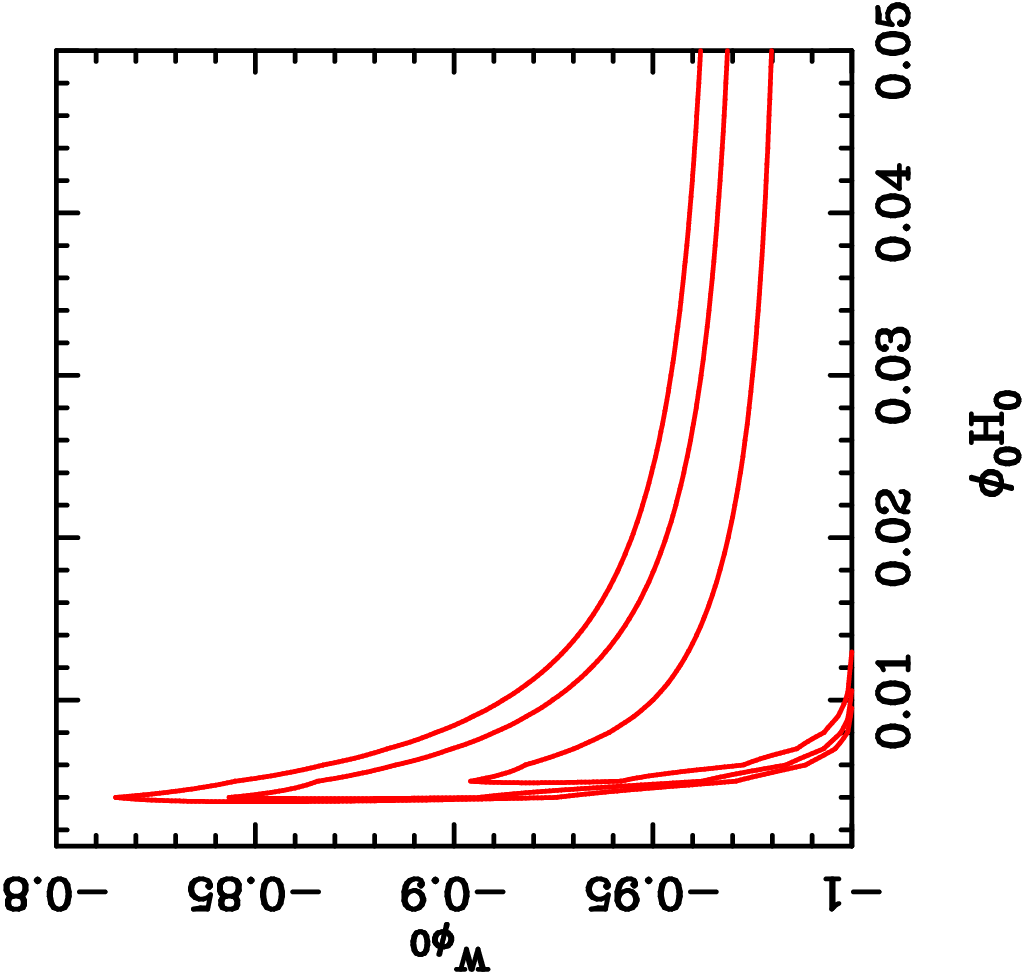}
    \includegraphics[width=0.33\textwidth,angle=-90]{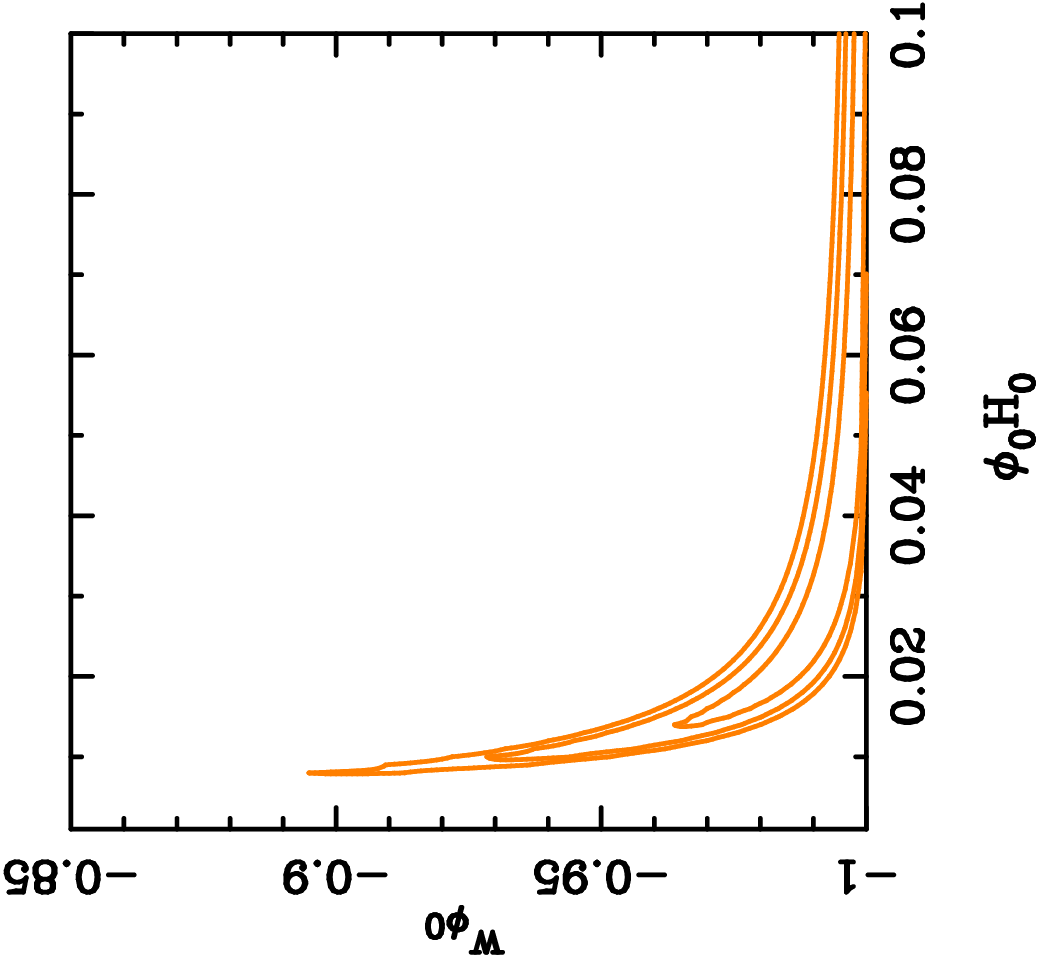}
    \includegraphics[width=0.33\textwidth,angle=-90]{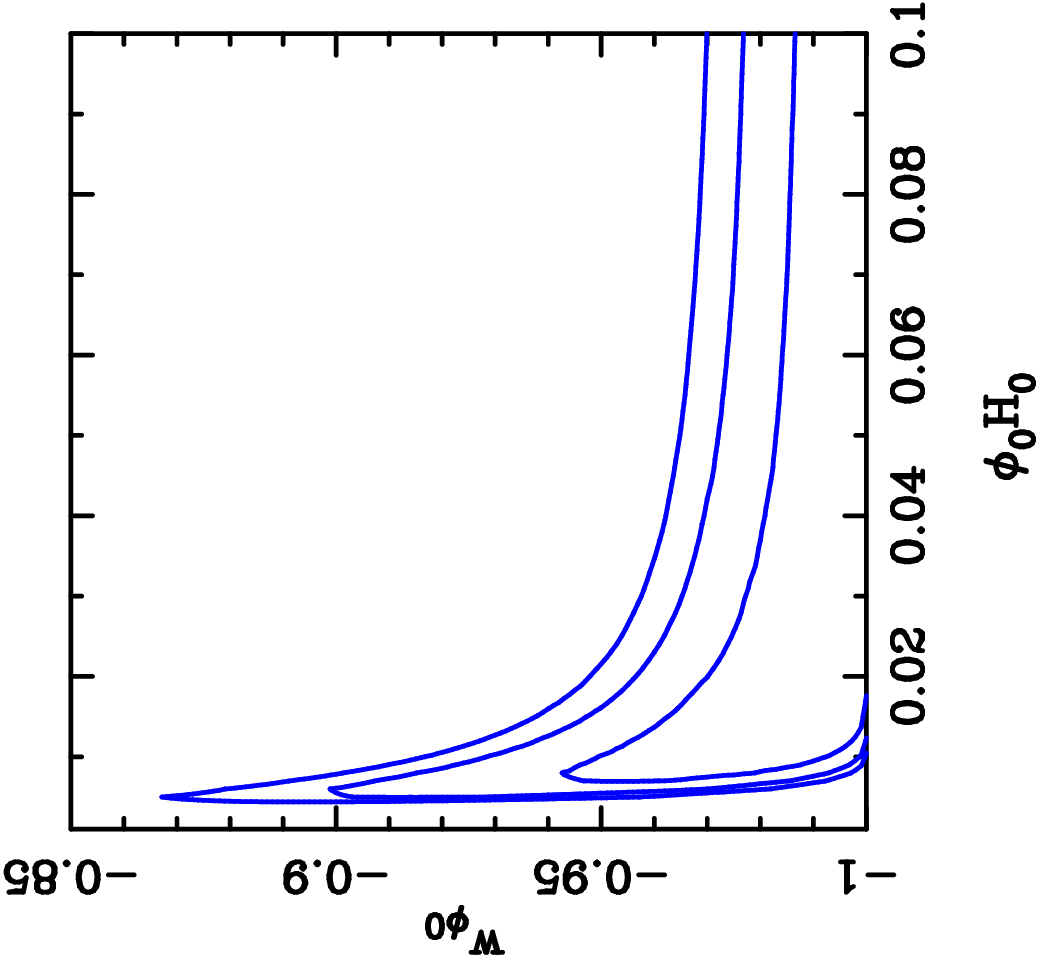}\\
    \includegraphics[width=0.33\textwidth,angle=-90]{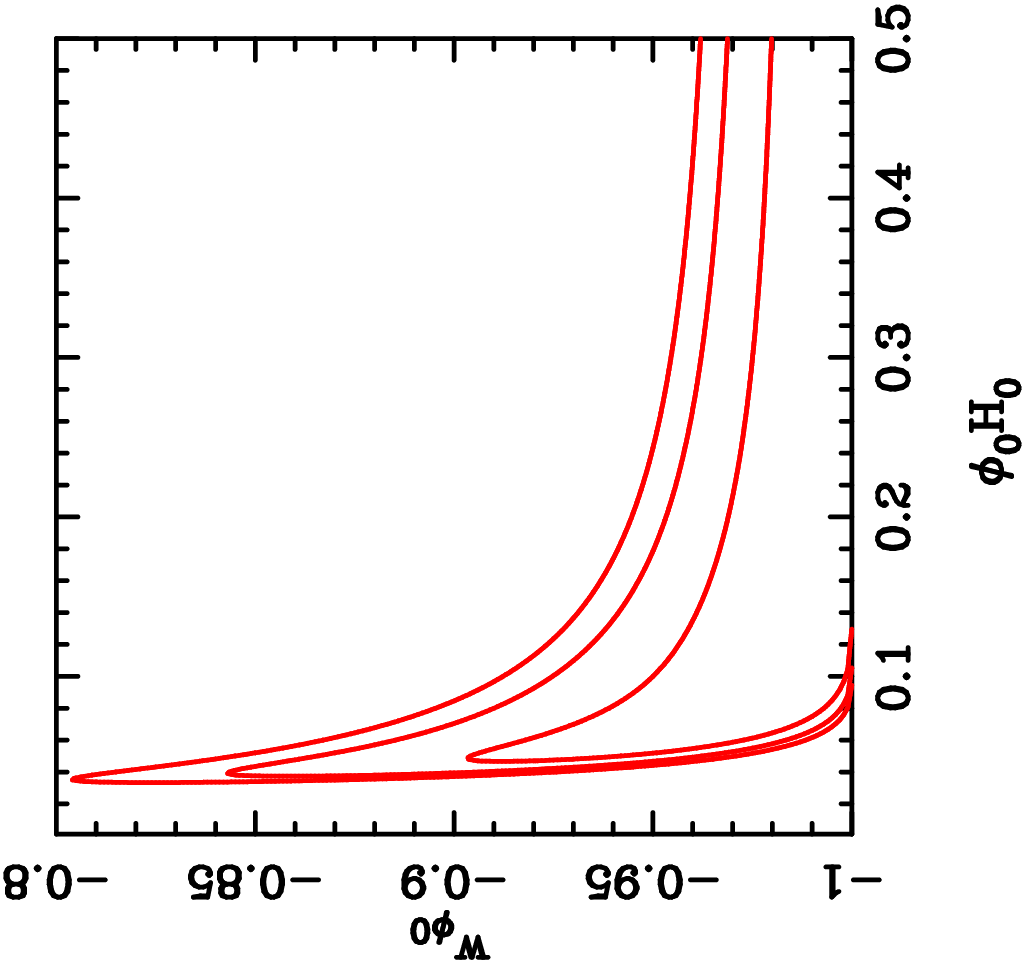}
    \includegraphics[width=0.33\textwidth,angle=-90]{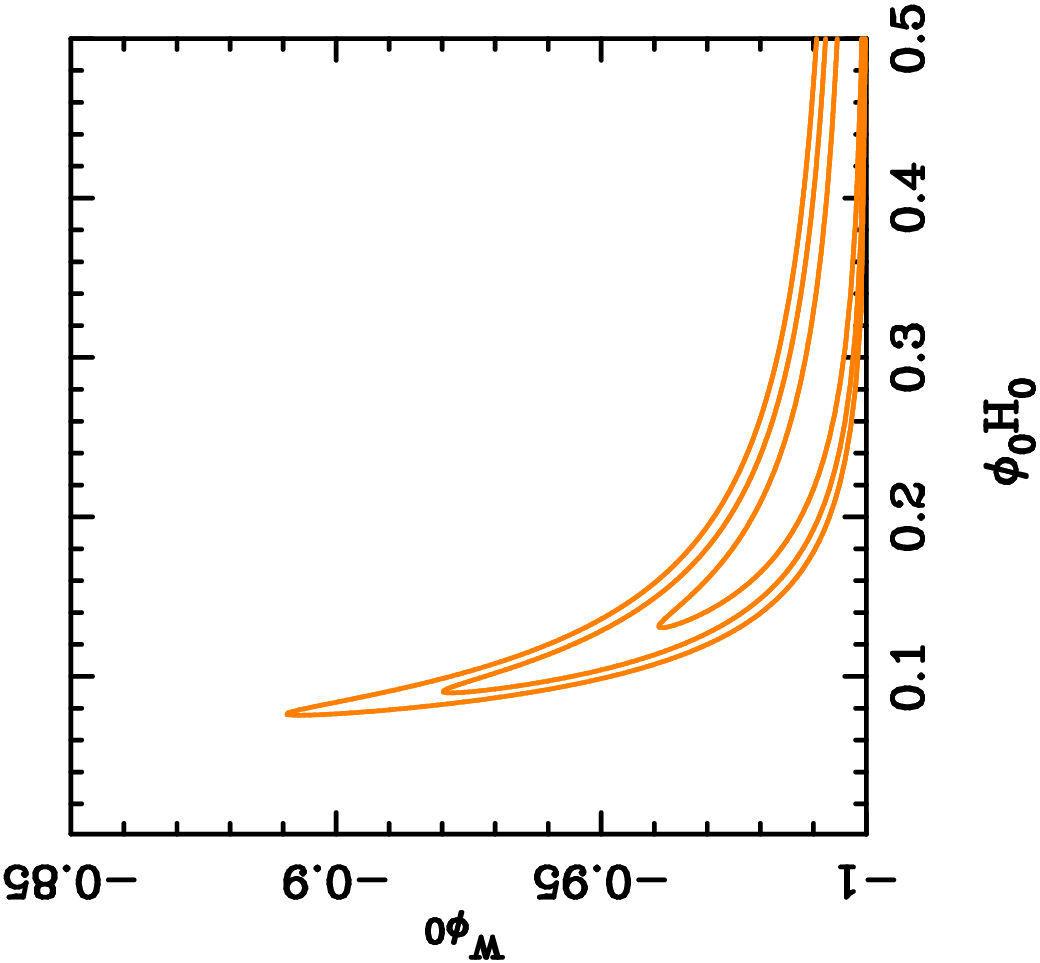}
    \includegraphics[width=0.33\textwidth,angle=-90]{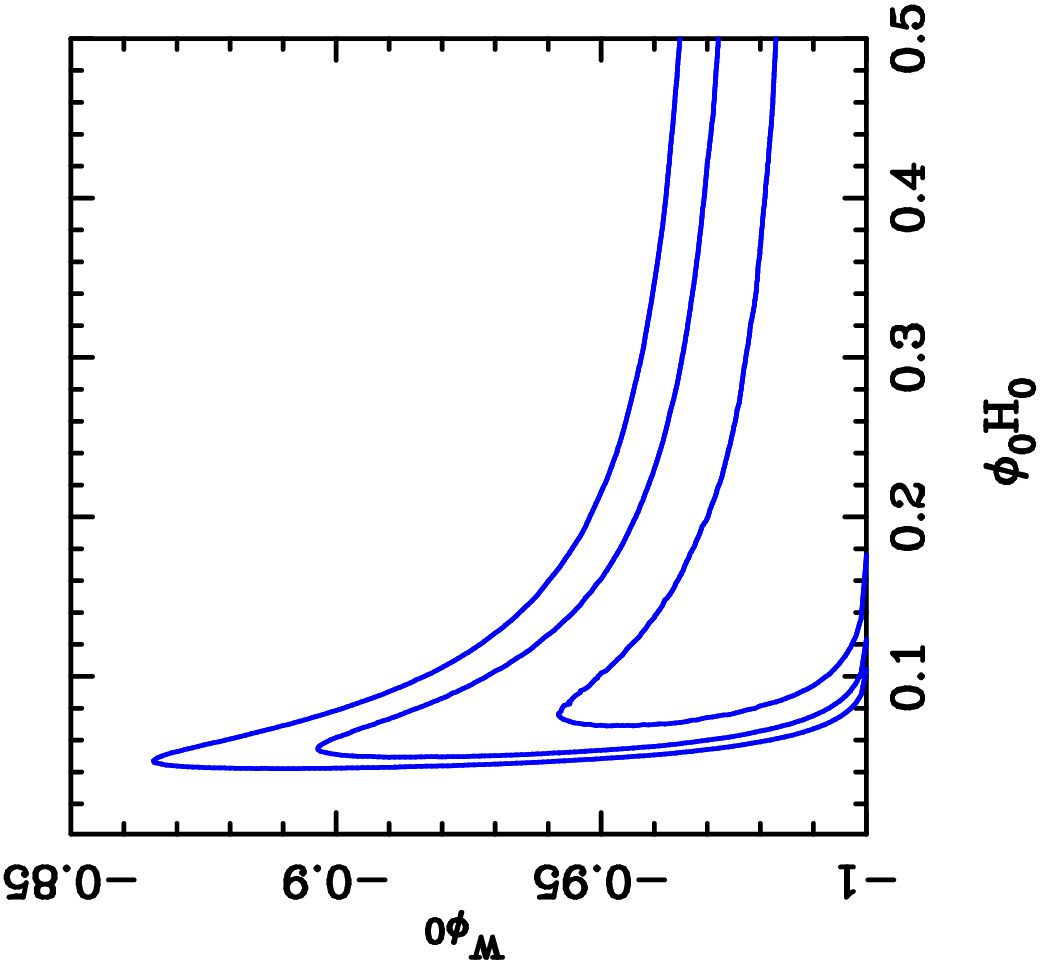}\\
    \includegraphics[width=0.33\textwidth,angle=-90]{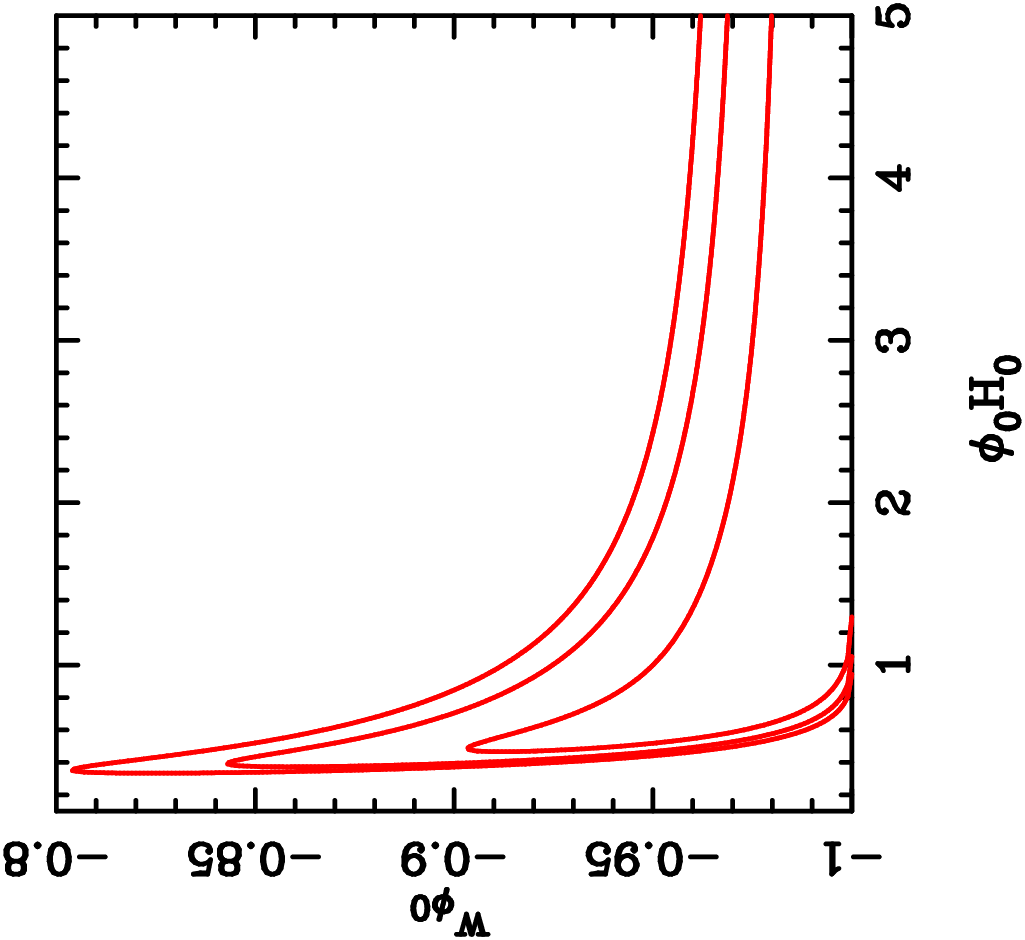}
    \includegraphics[width=0.33\textwidth,angle=-90]{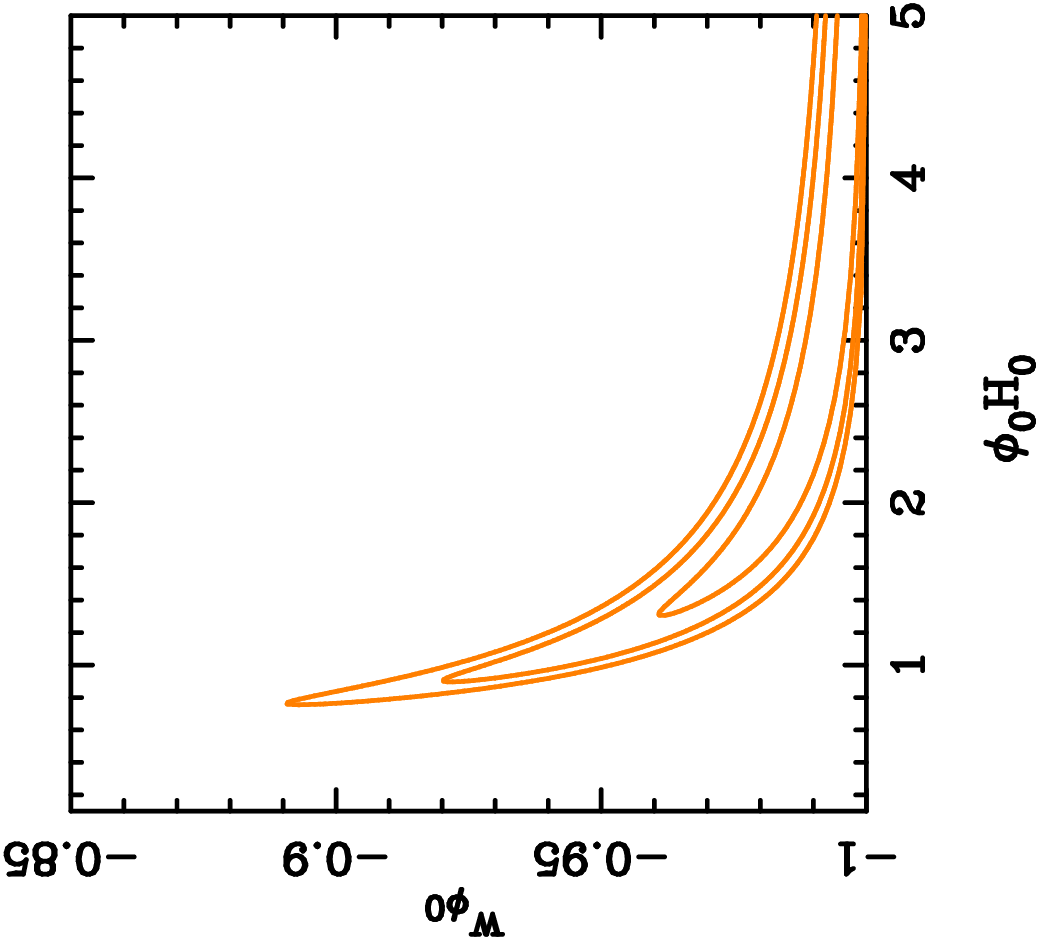}
    \includegraphics[width=0.33\textwidth,angle=-90]{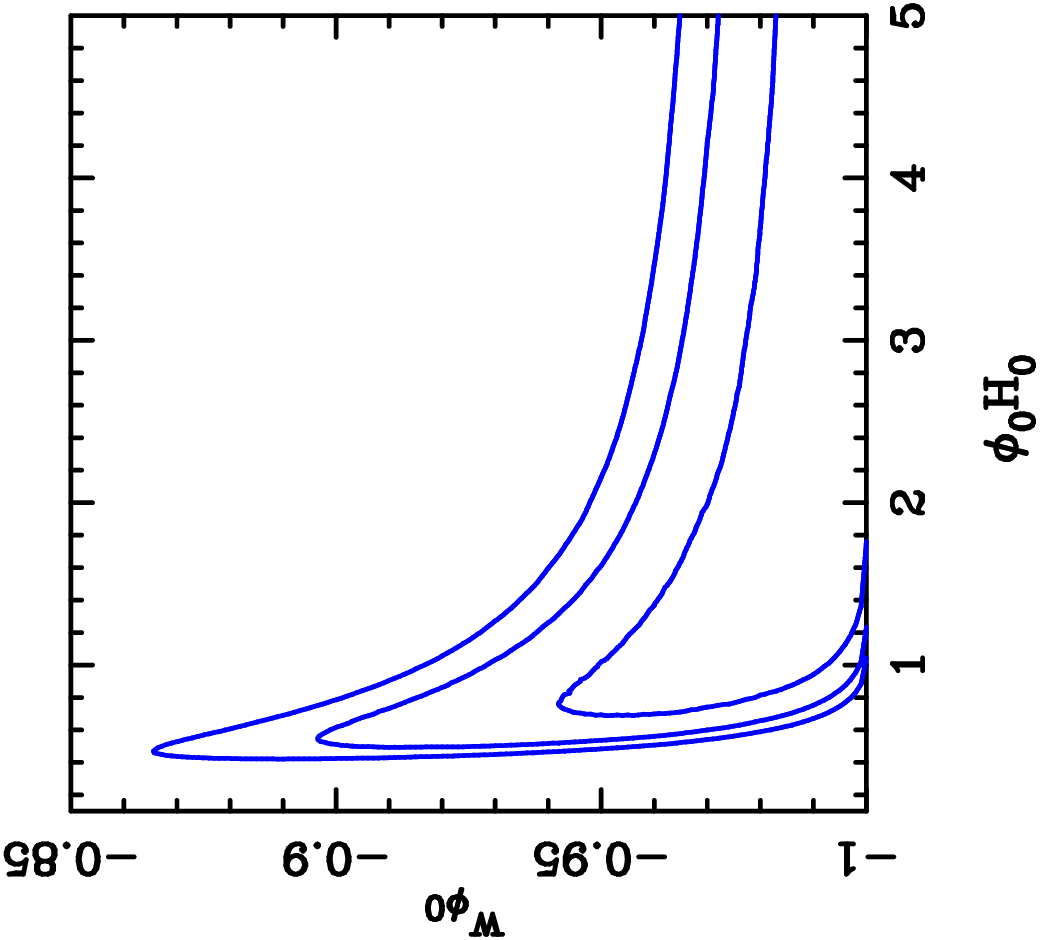}
    \caption{The figure shows we $1\sigma$, $2\sigma$ and $3\sigma$ confidence contours on
	    $w_{\phi 0}-\phi_0H_0$ plane for exponential potential. Red, orange and blue
	    colours represent BAO, Hz and SN-Ia data respectively.
	    First, second and third rows are for $\phi_0/\phi_a =~0.01,~0.1$ and $1.0$.
	    For all these plots we have set $\Omega_{m0}=0.285$. }
    \label{fig:expo_hph0VSwp}
\end{figure}

The evolution of the matter density parameter
$\Omega_m(z)$(red curves) and the dark energy
density parameter $\Omega_\phi(z)$(blue curves)
are shown in the plot on the right in figure~\ref{fig:evolution_invrse}.
We can see that even in the matter dominated era, dark energy contributes
significantly to the energy budget. 
For smaller values of $\phi_0H_0$, the contribution of dark energy, in the
matter dominated era, is larger than it is for larger value of this parameter. 
Here it should be noted that parameter $\phi_0H_0$ and $w_{\phi0}$ are
correlated. 
As we increase the value of $\phi_0H_0$ matter approaches complete
domination for large redshift.  

The evolution of the equation of state of the dark energy $w_\phi$ is shown
in the plot on the left in figure~\ref{fig:evolution_invrse}.
In the matter dominated era and before it, the equation of state of
tachyon dark energy is just like that of dust. 
After that, it starts evolving and make a sharp transition towards
smaller value than its present value $w_{\phi 0}$ then rises again. 
For a given value of $\phi_0H_0$ it maintains a constant value in the
future. 
For the larger values of $\phi_0H_0$, this constant value for future
evolution is closer to -1.0 as a larger value of $\phi_0H_0$ prefers
a cosmological constant like behaviour. 

\subsection{Constraints on Exponential Potential}
      \label{subsec:constraint_on_expo_pot}
      \begin{figure}[t]
  \includegraphics[width=0.33\textwidth,angle=-90]{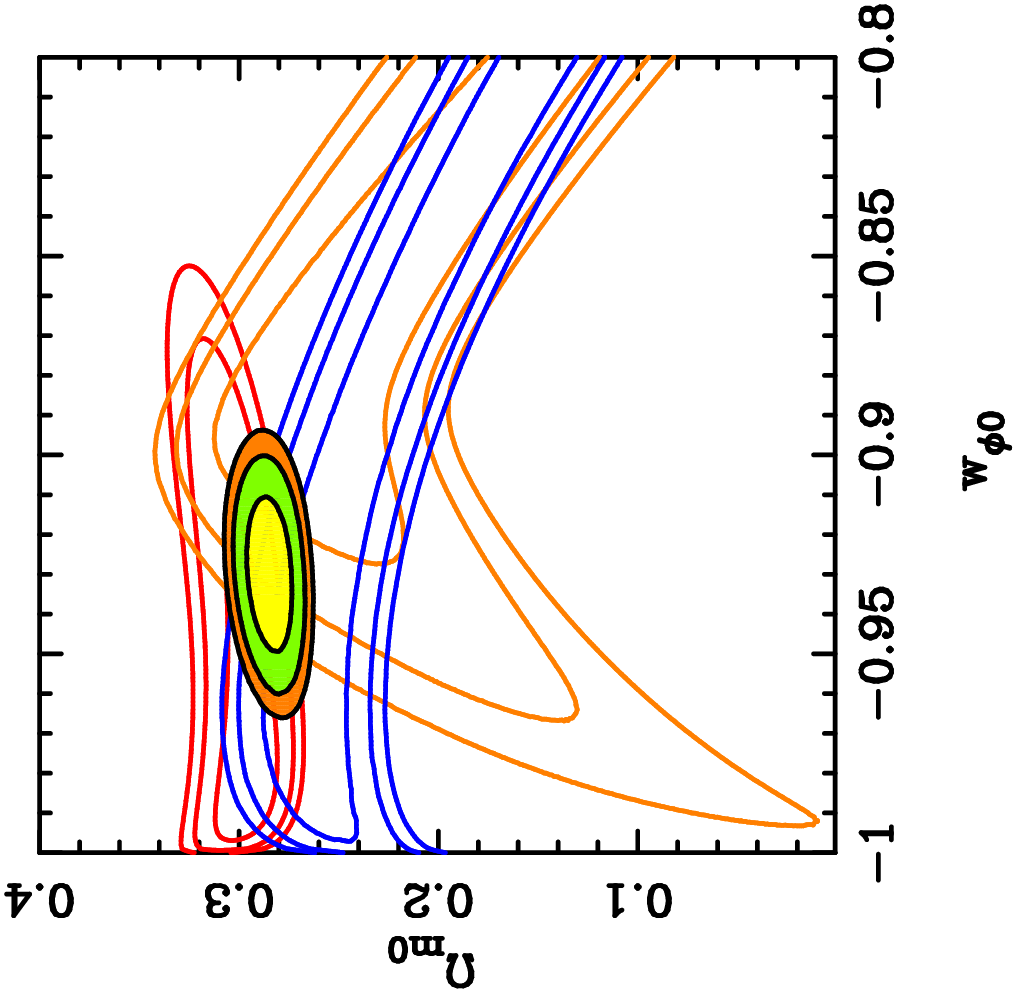}
  \includegraphics[width=0.33\textwidth,angle=-90]{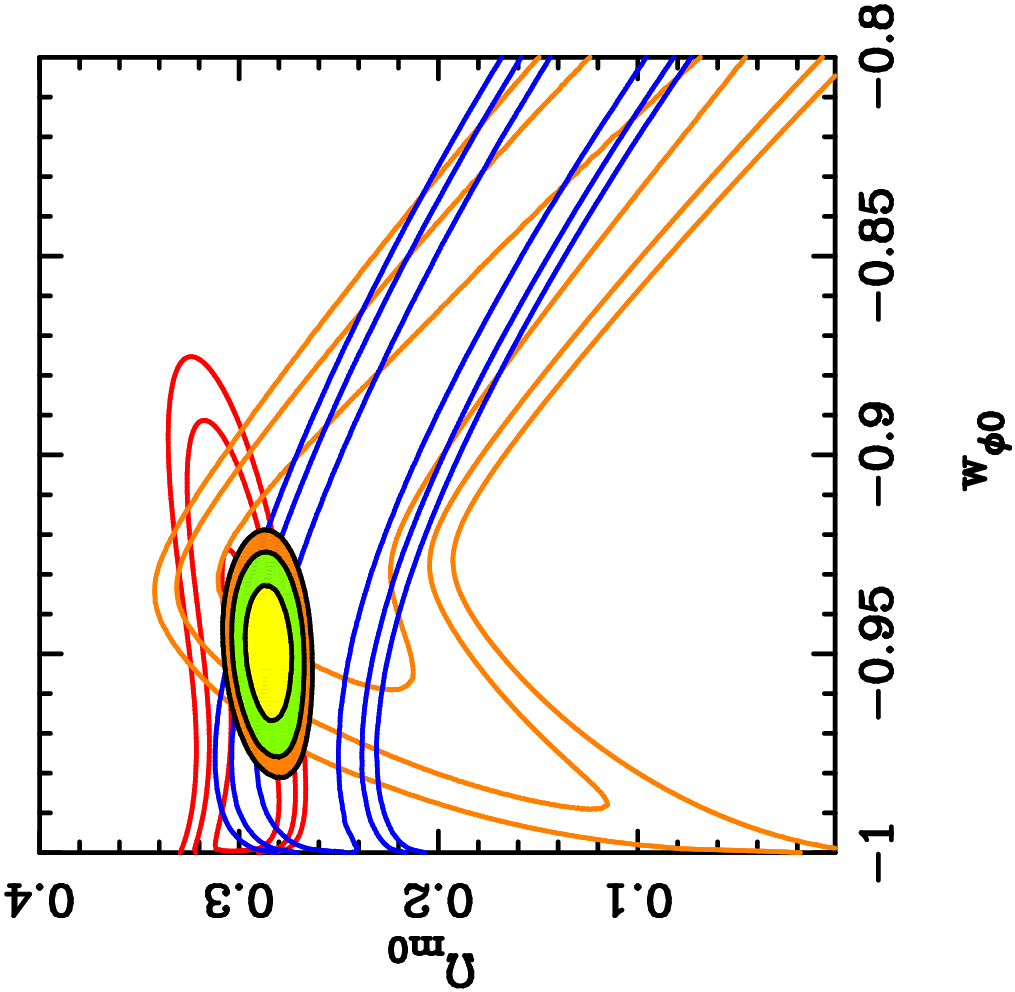}
  \includegraphics[width=0.33\textwidth,angle=-90]{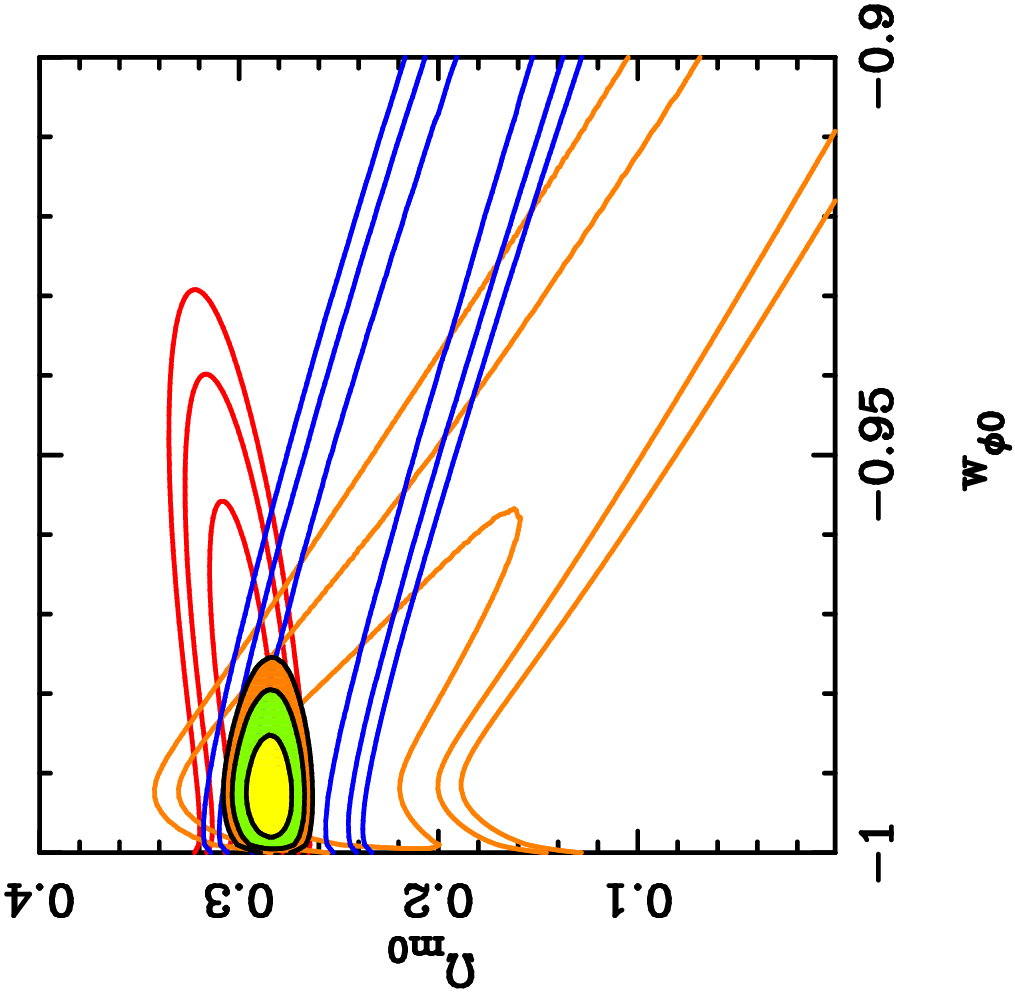}\\
  \includegraphics[width=0.33\textwidth,angle=-90]{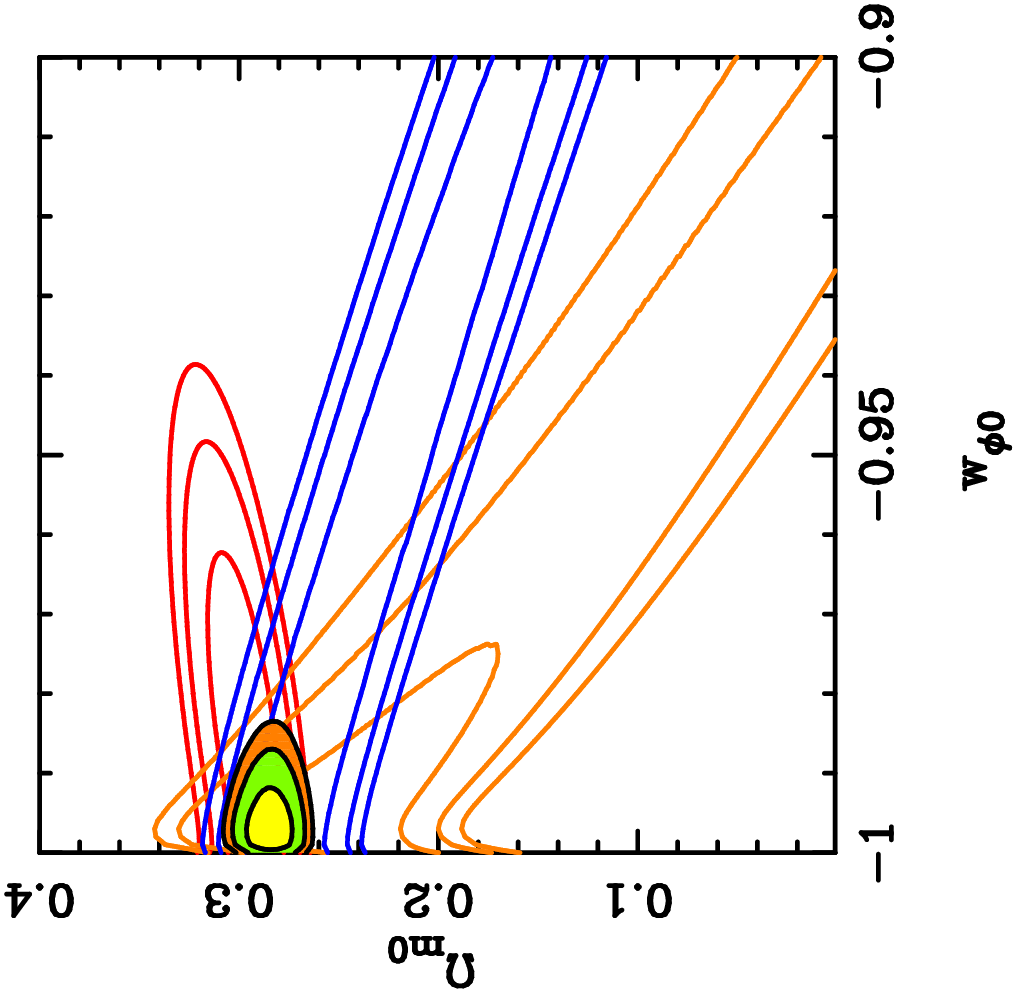}
  \includegraphics[width=0.33\textwidth,angle=-90]{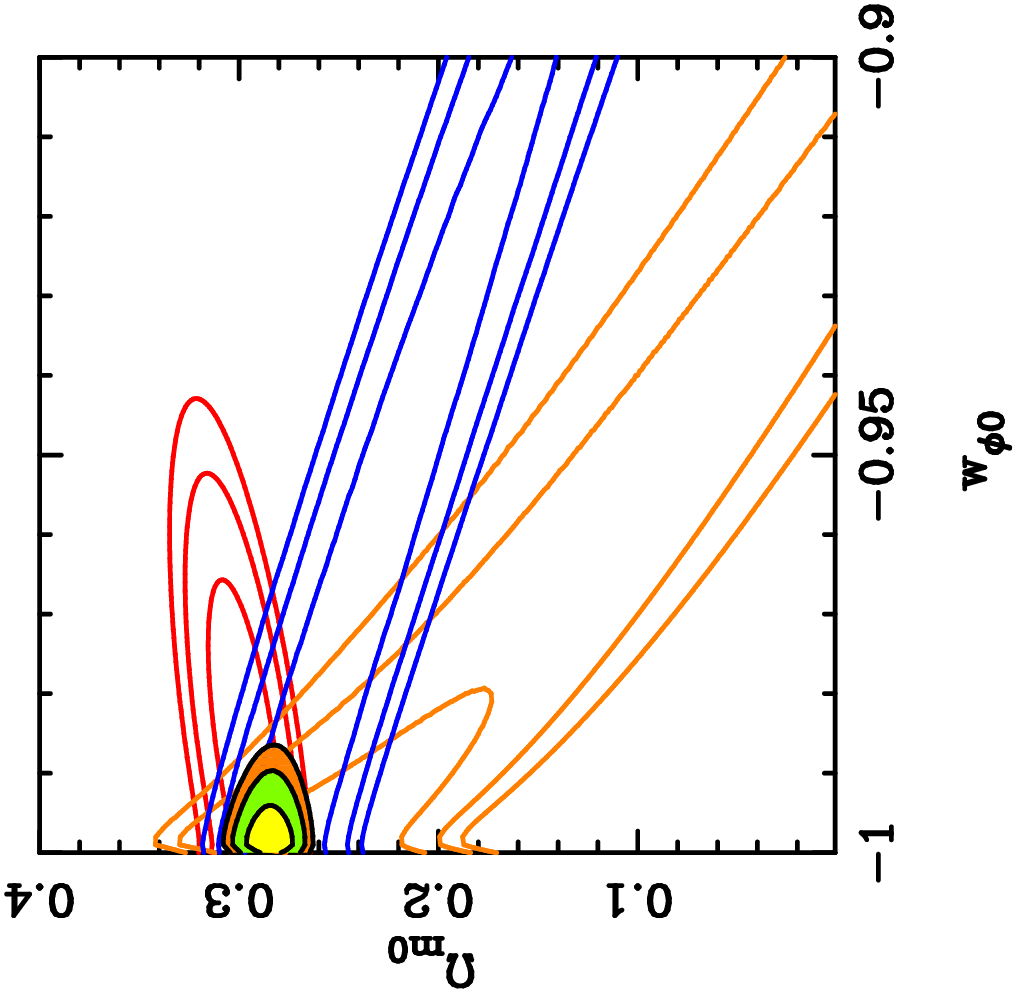}
  \includegraphics[width=0.33\textwidth,angle=-90]{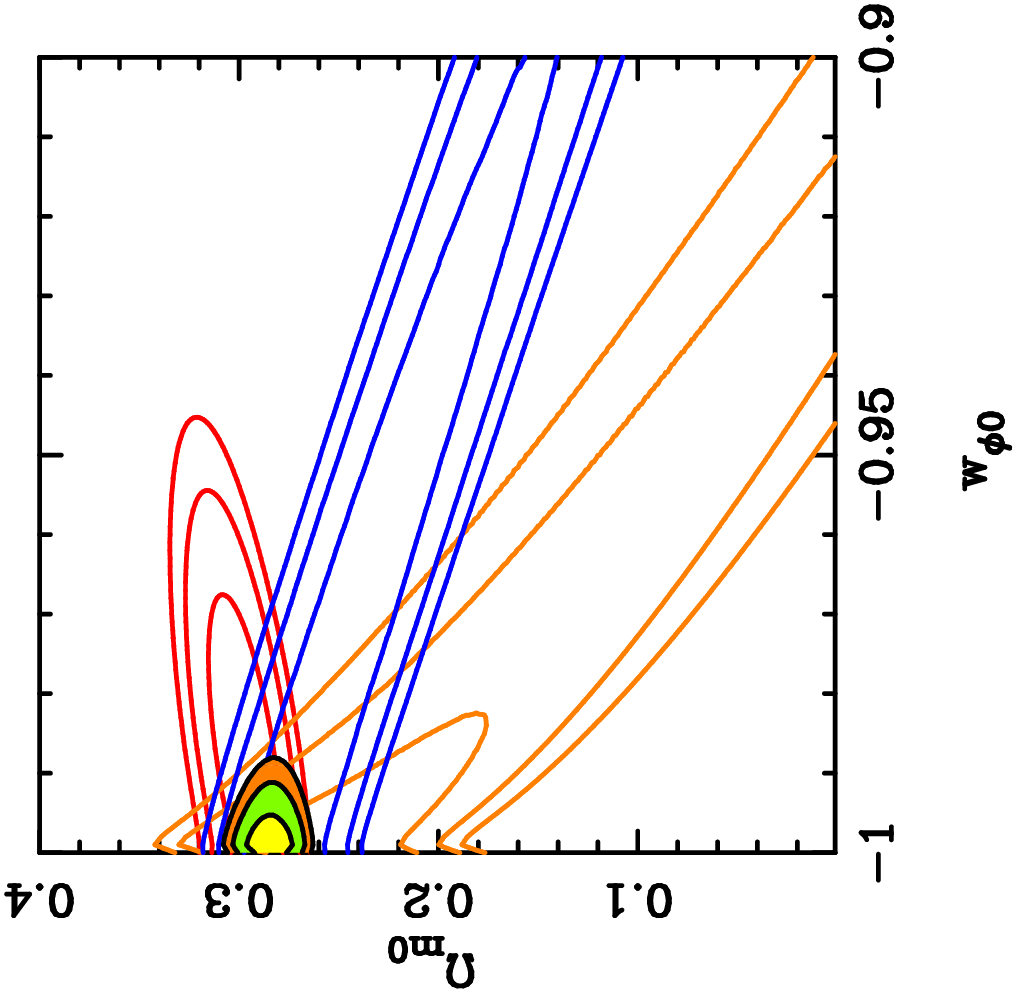}
  \caption{Contours in the $\Omega_{m0}$ - $w_{\phi 0} $ plane for a fixed value of
	   $\phi_0/\phi_a=0.1$ for the exponential potential. In the plots in row-1
	   $\phi_0H_0=~0.08,~0.1$ and $0.3$ from left to right and in row-2  $\phi_0H_0=
	   ~0.5,~0.7$ and $0.9$. The red, orange and blue colours represent BAO, 
	  H(z) and SN-Ia data respectively. Black, dark contours are for
	  combined dataset.}
  \label{fig:expo_omm_w}
\end{figure}
For the exponential potential, we need to constrain parameters
$\phi_0H_0, \phi_0/\phi_a, w_{\phi 0}$ and $\Omega_{m0}$. 
Rewriting the potential as $V=e^{ln(V_a)-\frac{\phi}{\phi_a}}$, we see that there
is an explicit degeneracy between $V_a$ and $\phi_{in}$, i.e., a change 
in $V_a$ and the corresponding change in $\phi_{in}$ leads to the same $V_{in}$.
Since we have replaced $V_a$ by the other parameters shown in
equation~(\ref{eq:amplitude_expo}), this degeneracy reflects in
degeneracy between $\phi_0/\phi_a$ and $\phi_0H_0$.

In figure~\ref{fig:expo_alldata} we show the agreement between data and theory with 
exponential potential~(\ref{eq:expo_potential}).
We plot the acoustic parameter $A(z)$, the Hubble parameter $H(z)$ and the distance modulus
$\mu(z)$ as a function of redshift $z$ along with the data points and the error bars,
taken from~\cite{Blake2011,Farooq2017,Suzuki2012}.
There are six overlapping theoretical curves in each of these plots representing the 
exponential potential.
To draw these theoretical curves, we have taken the best fit values of the parameters 
$\phi_0H_0$, $\Omega_{m0}$ and $w_{\phi0}$ from each row of table~\ref{tab:expo_omm_w}.   
\begin{table}[t]
        \centering
        \begin{tabular}{|c|c|c|c|c|}
        \hline
         &  &  & &  \\
        $\phi_0H_0$ & $\chi^2_{min}$ & $\Omega_{m0}$ & $w_{\phi 0}$ &  $\frac{V_a}{\rho_{cr}}$ \\
         &  &  & &  \\
        \hline
         &  &  & &  \\
        0.08 & 600.125 &  $0.285^{+0.022}_{-0.022}$&  $-0.928^{+0.034}_{-0.038}$ & [0.724,0.801]  \\
         &  &  & &  \\
        \hline
        &  &  & & \\
        0.10 &  595.862 & $0.285^{+0.023}_{-0.022}$ & $-0.949^{+0.030}_{-0.032}$& [0.733,0.806]  \\
         &  &  & &  \\
        \hline
        &  &  & &  \\
        0.30 & 590.327 & $0.285^{+0.023}_{-0.022}$ &$-0.993^{+0.017}_{-0.007}$  & [0.756,0.815] \\
        &  &  & &  \\
        \hline
        &  &  & &  \\
        0.50 & 590.136 & $0.284^{+0.024}_{-0.021}$ & $-0.997^{+0.013}_{-0.003}$& [0.759,0.815]\\
        &  &  & &   \\
        \hline
         &  &  & &   \\
         0.70 & 590.132 & $0.285^{+0.023}_{-0.022}$ & $-0.999^{+0.012}_{-0.001}$ & [0.760,0.815] \\
         &  &  & &   \\
        \hline
        &  &  & &   \\
         0.90 & 590.061 & $0.285^{+0.023}_{-0.022}$ & $-0.999^{+0.011}_{-0.001}$ & [0.760,0.815] \\
         &  &  & &   \\
        \hline
        \end{tabular}
        \caption{Best fit values for $\Omega_{m0}$ and $w_{\phi 0}$ with $3\sigma$ confidence interval
		for the exponential potential for different values of $\phi_0H_0$ for combine data (BAO +
		H(z) + SN-Ia) set. Here we have fixed the value of $\phi_0/\phi_a = 0.1 $. In the last 
		column, we have shown the range of amplitude of potential $V_a$ normalized by present
		critical density $\rho_{cr}$. It is calculated from equation~(\ref{eq:amplitude_expo})
		considering $3\sigma$ confidence range of $\Omega_{m0}$ and $w_{\phi 0}$.}
        \label{tab:expo_omm_w}
    \end{table}
    
To show the degeneracy mentioned above, we first plot the $1\sigma$, $2\sigma $ and
$3\sigma$ contours in  $w_{\phi 0} -\phi_0H_0$ plane in figure~\ref{fig:expo_hph0VSwp}. 
In these plots, we have fixed $\Omega_{m0} = 0.285 $ and first, second
and third row are for $\phi_0/\phi_a=0.01,~0.1$ and $1.0$.
We can see that all datasets (BAO, Hz and SN-Ia) have lower bound on
$\phi_0H_0$. 
The lower bound on this parameter depends on the value of parameter
$\phi_0/\phi_a$; for $\phi_0/\phi_a = 0.01 $ we have $\phi_0H_0
\gtrsim 4\times 10^{-3} $, for  $\phi_0/\phi_a = 0.1$ we have $\phi_0H_0 \gtrsim
0.04 $ and $\phi_0/\phi_a = 1.0$ we have $\phi_0H_0 \gtrsim 0.41 $.
The lower bound on $\phi_0H_0 $ increases with $\phi_0/\phi_a $. 
We fix the parameter $\phi_0/\phi_a = 0.1$, and we do our analysis by
keeping other parameters free. 
The analysis below is equally valid for any other value of this parameter 
if $\phi_0H_0$ is adjusted accordingly or properly scaled.  
\begin{figure}[t]
    \includegraphics[width=0.35\textwidth,angle=-90]{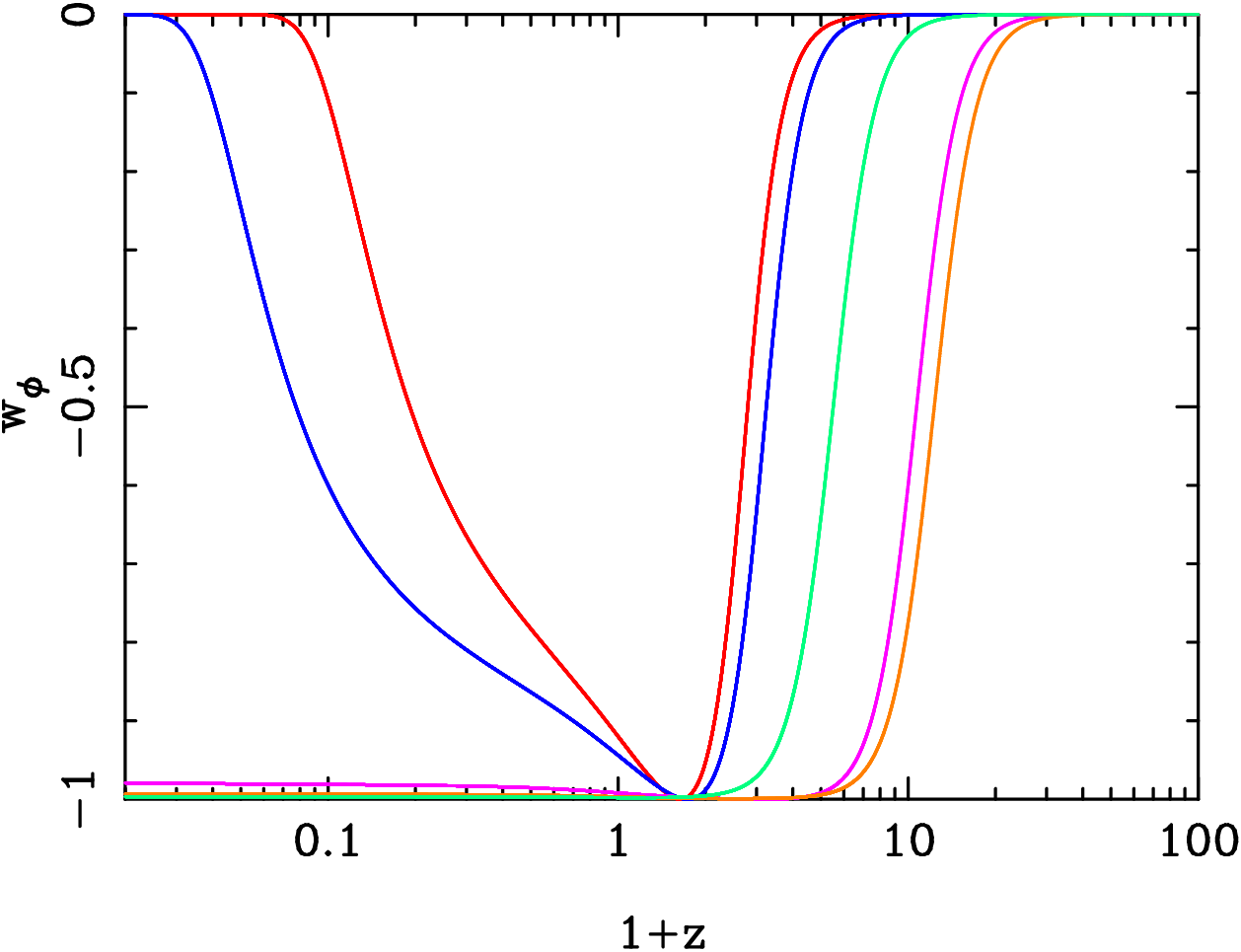}
    \includegraphics[width=0.35\textwidth,angle=-90]{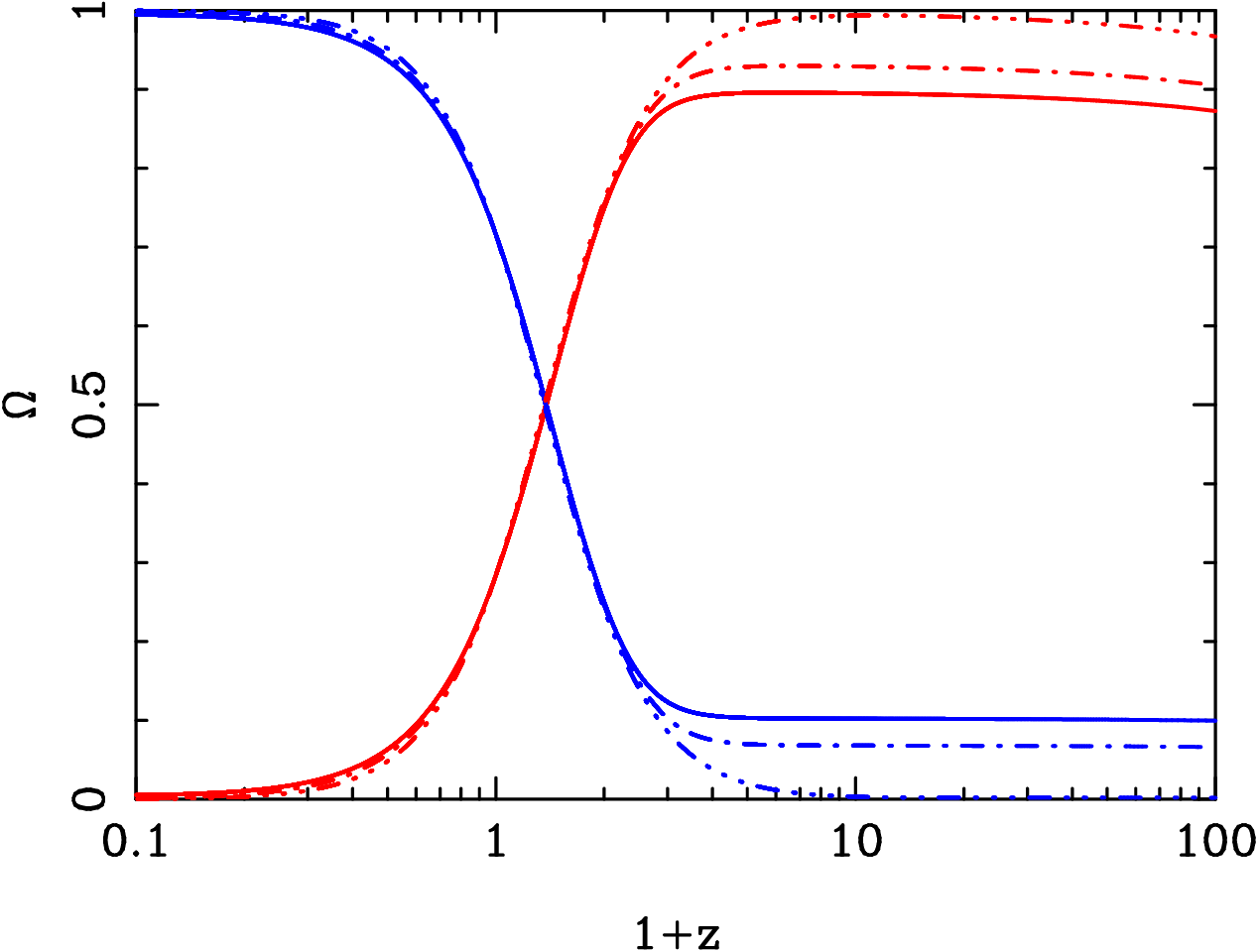}
    \caption{The plot on the left shows the evolution of equation of state $w_\phi$ with
	     redshift for exponential potential. Red, blue, sky blue, orange and gray
	     lines represent $\phi_0H_0=~0.08,~0.1,~0.3,~0.5$ and $0.7$.
	     In the plot on the right evolution of $\Omega_m $ (red curves) and
	     $ \Omega_\phi $ (blue curves) with redshift are shown.
	     Solid, dash-dot and dashed-dot-dot-dot lines represent $\phi_0H_0=~0.08,~0.1$ and $0.3$.
	     Parameter $\Omega_{m0}$ and $w_{\phi 0}$ are the best fit values taken from 
	     table~\ref{tab:expo_omm_w} for the corresponding value of $\phi_0H_0$.
	     Parameter $\phi_0/\phi_a = 0.1$.}
    \label{fig:evolution_expo}
\end{figure}

We have shown constraints on $\Omega_{m0} - w_{\phi 0}$ plane
in figure~\ref{fig:expo_omm_w} for $\phi_0H_0 = 0.08,~0.1,~0.3,~0.5,
~0.7 $ and $0.9 $. 
Here we have fixed $\phi_0/\phi_a = 0.1$.
The contours filled with three different colours represent result for
the combination of datasets. 
The $3\sigma$ results for this model are shown in table~\ref{tab:expo_omm_w}.  
We started from $\phi_0H_0=0.08$ as for smaller values, the value of
$\chi^2_{min}$  increases sharply.
We can see that for a smaller value of $\phi_0H_0$ the three datasets
are not in good agreement  with each other and hence a large value of
$\chi^2_{min}$. 
As we increase the value of this parameter, the value of
$\chi^2_{min}$ decreases and the combined contours become smaller.
The BAO data provides the tightest constraint on $\Omega_{m0}$ among all; this is
consistent with previous studies~\cite{Tripathi:2016slv,Sangwan:2017kxi}.
The value of this parameter is $\Omega_{m0}=0.285^{+0.023}_{-0.022}$
with $3\sigma$ confidence for the combination of all three datasets, and it is almost a
constant with variation in $\phi_0H_0$. 
On the other hand, the value of $w_{\phi 0}$ depends on $\phi_0H_0$.
In the last column of table~\ref{tab:expo_omm_w}, we have shown the
$3\sigma$ allowed range of the amplitude of the potential normalized
to the present day value of the critical density $\rho_{cr} =
\frac{3H^2_0}{8\pi G}$ using equation~(\ref{eq:amplitude_expo}),
considering $3\sigma$ confidence interval of $\Omega_{m0}$ and
$w_{\phi 0}$. 
From equation~(\ref{eq:amplitude_expo}) it is clear that the amplitude
of the potential is not explicitly dependent on $\phi_0H_0$ and as we have fixed
$\phi_0/\phi_a = 0.1$; its value only depends on other parameters
$\Omega_{m0}$ and $ w_{\phi 0}$. 
Since the values of these parameters saturates with an increase in
$\phi_0H_0$, the amplitude of potential also approaches a fixed
value unlike the case of tachyon model with inverse square potential.

The evolution of the equation of state parameter at different epochs
in the expansion history of the universe 
is shown in the plot on the left in figure~\ref{fig:evolution_expo}.
For a tachyon field with an exponential potential, the accelerating phase
is sandwiched between two decelerating phases. 
In future, the universe goes back to a decelerating phase and duration
of the accelerating phase depends on the value of $\phi_0H_0$ and $w_{\phi 0}$.
In this plot, we can see that for a smaller value of
$\phi_0H_0$, this duration is small and the universe goes to decelerating phase
once again in relatively near future than it is for larger values of
this parameter. 
The notable thing here is that parameters $\phi_0H_0$ and $w_{\phi 0}$ are
correlated and for small $\phi_0H_0$ the best fit value of $w_{\phi0}$
is large or away from -1.  

We can see that in the matter dominated era, the dark energy behaves like a
fluid and in the near past, it starts to deviate from $w_\phi=0$ sharply.
For a larger value of $\phi_0H_0$, its deviation begins earlier.
At first, it goes close to $-1$ depending on its present day value
$w_{\phi0}$ and then it rises away from $-1$. 
For a smaller value of $\phi_0H_0$ it faster approach to a fluid like
equation of state  $w_\phi=0$ and as it crosses the condition $w_\phi \le
-1/3$ for an accelerated expansion and the universe goes to a
decelerating phase once again.  
We have shown the evolution of density parameters
$\Omega_m $(red curves) and $ \Omega_\phi $(blue curves) with redshift
in the plot on the right in figure~\ref{fig:evolution_expo}.
In the matter dominated era, matter does not fully dominate the energy budget.
Part of sub-dominated dark energy density parameter is large (solid line)
for a smaller value of $\phi_0H_0$ and as we increase the value of this parameter non-relativistic
matter dominates the energy of the universe completely.  
\section{Summary and Conclusions}
      \label{sec::SummaryandConclusions}
 In this work, we have constrained parameters of the tachyon dark
 energy model with an inverse  square potential and an exponential
 potential. 
 For this purpose, we have used the Baryon acoustic Oscillation data
 (from SDSS DR12, 6dFGS, SDSS DR7, WingleZ surveys), direct  measurement of
 Hubble parameter (H(z)) data and Supernova-Ia Union 2.1 data.   
 For the inverse square potential, we have three parameters $\phi_0H_0$,
  $w_{\phi 0}$ and $\Omega_{m0}$. 
 For the exponential potential, apart from these three, an extra
 parameter $\phi_0/\phi_a$ is present.  
 There is a lower bound on the parameter $\phi_0H_0$, and all larger
 values are allowed. 
 For the inverse square potential, $\phi_0H_0 \ge 0.775$ at the $3\sigma$
 confidence level.
 For the exponential potential, this value depends on $\phi_0/\phi_a$,
 and the lower bound on  $\phi_0H_0$ increases with increase in
 $\phi_0/\phi_a$. 
 Using combined data of all three measurements, we find that the present day
 matter density  parameter is constrained to the values
 $\Omega_{m0}=0.285^{+0.023}_{-0.022}$ at the $3\sigma$ confidence for
 both the potentials and it remains almost same with  variation in
 $\phi_0H_0$.
This value of $\Omega_{m0}$ for the tachyon model is less than
	 the value of this parameter for a flat $\Lambda CDM$ model as determined by  
	 current observations, e.g., $\Omega_{m0}=0.295\pm 0.034$ ( at $68\%$ confidence
	 using JLA data ~\cite{Betoule2014}), $\Omega_{m0}=0.3089\pm 0.0062$
	 (at $68\%$ confidence obtained from CMB-TT,TE,EE+ low-P + lensing+ BAO+JLA+ 
	 $H_0$ data ~\cite{Planck-2015}), $\Omega_{m0}=0.311\pm 0.0056$
	 (at $68\%$ confidence obtained from  CMB-TT,TE,EE+ low-P + lensing+ BAO data
	 ~\cite{Planck-2018})  and $\Omega_{m0}=0.310\pm 0.005 $ (at $95\%$ confidence
	 using BAO DR12 + SN-Ia data~\cite{Alam2017}).
	 The value of $\Omega_{m0}$ for tachyon model is in agreement with its
	 value  for flat $\Lambda CDM$ model constrained by the JLA data  within  $1\sigma$.
	 There is a tension with constraints from Planck and BAO DR12 data.
 
 The value of $w_{\phi 0}$ depends on  $\phi_0H_0$. 
 For a smaller value of $\phi_0H_0$, the equation of state parameter $w_{\phi 0}$ 
 has a larger value and as its value increases, $w_{\phi 0}$
 approaches $-1$.
 A large range of $\phi_0H_0$ is allowed by the background data.
 The parameter $\phi_0H_0$ need to be tuned to obtain the value of 
 the equation of state parameter $w_{\phi 0}$ which is supported by
 observations ($w_{\phi 0}=-1.006 \pm 0.045$~\cite{Planck-2015} and
 $w_{\phi 0}=-1.03 \pm 0.03$~\cite{Planck-2018} ). 
 This tuning is not as severe as the fine-tuning
 problems in $\Lambda CDM$ model.
 This parameter is constrained from below to a value closer to unity, and 
 there is no upper bound. Therefore the tuning of this parameter is not severe.
 The potentials, we have used in this paper, have also been
	 	 extensively used for canonical scalar field (quintessence field) model
	 	 of dark energy and similar results have been
	 	 found~\cite{Ratra:1987rm,Linder:2007wa,MacorraStephan-Otto2002,
	 	 Zlatev_etal1999,RiazueloUzan2000,MatosArturo2001}.
		 Specially, tracker solutions of quintessence model are able to solve the
		 fine-tuning  problem, and  thawing or freezing model ameliorate this
		 problem~\cite{Ratra:1987rm,Linder:2007wa,Zlatev_etal1999,MatosArturo2001}. 
		 In~\cite{MacorraStephan-Otto2002}, it has been shown that for the potential
		 $V(\phi)\propto \phi^{-n}$, with $n<5$, the solutions do not have fine-tuning
		 problem and a large range of initial conditions provide acceptable solutions.
		 Similar results have also been shown in~\cite{Ratra:1987rm,Linder:2007wa,
		 Zlatev_etal1999,Anjan_etal2012} for inverse power law potentials.
		 Exponential potential have been studied in~\cite{Ratra:1987rm,FerreiraJoyce1997,
		 Wetterich1995,Anjan_etal2012} for quintessence model and it is found to ameliorate
		 the fine-tuning problem.
		 In our study, tachyon models with both the potentials generate acceptable solution
		 for large range of parameters.	 
		 On the other hand, for larger value of $\phi_0H_0$, it is able to mimic the
		 cosmological constant like equation of state at present.
		 Hence, in the light of current observational data, tachyon model is an interesting and
		 important alternate for dark energy.
 
 We have also studied the evolution of the phases of expansion, the density
 parameters and the equation of state of dark energy with redshift. 
 We find the transition redshift to be in the range $0.61 \lesssim
 z_{ac} \lesssim 0.80$.
 For the exponential potential, the duration of the acceleration phase
 depends on $\phi_0H_0$ and $w_{\phi 0}$ (as these parameters are
 correlated). 
 For a smaller value of $\phi_0H_0$ this duration is small.
 The equation of state of the tachyon dark energy, in the matter dominated and
 earlier phases, is dust like ($w_{\phi}=0$).
 It then makes a sharp transition to that of a cosmological constant
 as dark energy domination begin.
 The value of the equation of state parameter  rises again to match
 its present day value $w_{\phi 0}$. 
 For the inverse square potential, it approached a  constant  value
 depending on the values of  $\phi_0H_0$ and $w_{\phi 0}$.
 For exponential potential, it rises towards $w_{\phi}=0$, and as it
 becomes greater than $-1/3$, the  universe once again goes to a
 decelerating expansion  phase.
 For tachyon dark energy, matter does not fully dominate the energy budget.
 However, as we increase the value of parameter $\phi_0H_0$, it approaches
 to  dominating fully as the equation of state approaches like that of a
 cosmological constant.

 The constraints obtained here are stringent, and there is a clear
 preference for models which are close to the cosmological constant model.
 A specific set of parameters can be ruled out in a  given set of
 models whereas current data cannot completely distinguish between different
 models and does not fully rule out any.
 The range of the combined constraints on the matter density parameter
 and the equation of state parameter are determined largely by the BAO
 data and by the supernova data respectively.
 While the Hubble parameter data constrains the parameters well, the
 allowed range is larger than that allowed by  other observations.
 This is possibly due to the fact that the Hubble parameter
 measurement data is a compilation of measurements with different
 methods and accompanies measurements of different cosmological
 quantities. 
The constraints on the parameters are stringent and more data, and
further studies in perturbations in tachyon dark energy  are likely to
break the  degeneracy between different models which are 
 allowed  by pure distance  measurements.
 
 \section{Acknowledgements}
 The numerical work in this paper was done using the High Performance Computing facility at IISER Mohali.
 The authors thank J. S. Bagla and Ankan Mukherjee for helpful discussions.



\bibliography{referencesBGtachyon}
 
\bibliographystyle{JHEP}

\printindex
\end{document}